\documentclass[manuscript]{aastex}
\usepackage{amsmath}
\usepackage{graphicx}
\usepackage{lscape}
\usepackage{color}

\makeatletter 
\def\appendix{
\def\thetable{\Alph{section}\arabic{table}} 
\setcounter{table}{0}
\def\thesection{\Alph{section}} 
\setcounter{section}{0} 
\def\@seccntformat##1{ 
\@nameuse{prefix@##1} 
\@nameuse{the##1}
\@nameuse{postfix@##1}\quad} 
\def\prefix@section{Appendix~} 
\def\postfix@section{:}
} 
\makeatother 

\shorttitle{Chemistry in the First Hydrostatic Core Stage}
\shortauthors{Furuya et al.}

\begin{document}


\title{CHEMISTRY IN THE FIRST HYDROSTATIC CORE STAGE BY ADOPTING THREE-DIMENSIONAL RADIATION HYDRODYNAMIC SIMULATIONS}


\author{Kenji Furuya\altaffilmark{1}, Yuri Aikawa\altaffilmark{1}, Kengo Tomida\altaffilmark{2,3,4}, Tomoaki Matsumoto\altaffilmark{5}, Kazuya Saigo\altaffilmark{3}, Kohji Tomisaka\altaffilmark{2,3}, Franck Hersant\altaffilmark{6,7}, and Valentine Wakelam\altaffilmark{6,7}}
\affil{\altaffilmark{1}Department of Earth and Planetary Sciences, Kobe University, Kobe 657-8501, Japan}
\affil{\altaffilmark{2}Department of Astronomical Science, The Graduate University for Advanced Studies (SOKENDAI), Osawa, Mitaka, Tokyo 181-8588, Japan}
\affil{\altaffilmark{3}National Astronomical Observatory of Japan, Mitaka, Tokyo 181-8588, Japan}
\affil{\altaffilmark{4}Department of Astrophysical Sciences, Princeton University, Princeton, NJ 08544, USA}
\affil{\altaffilmark{5}Faculty of Humanity and Environment, Hosei University, Fujimi, Chiyoda-ku, Tokyo 102-8160, Japan}
\affil{\altaffilmark{6}Univ. Bordeaux, LAB, UMR 5804, F-33270, Floirac, France}
\affil{\altaffilmark{7}CNRS, LAB, UMR 5804, F-33270, Floirac, France}
\email{furuya@stu.kobe-u.ac.jp}




\begin{abstract}
We investigate molecular evolution from a molecular cloud core to a first hydrostatic core in three spatial dimensions. We perform a radiation hydrodynamic simulation in order to trace fluid parcels, in which molecular evolution is investigated, using a gas-phase and grain-surface chemical reaction network. We derive spatial distributions of molecular abundances and column densities in the core harboring the first core. We find that the total of gas and ice abundances of many species in a cold era (10 K) remain unaltered until the temperature reaches $\sim$500 K. The gas abundances in the warm envelope and the outer layer of the first core ($T \lesssim 500$ K) are mainly determined via the sublimation of ice-mantle species. Above 500 K, the abundant molecules, such as H$_2$CO, start to be destroyed, and simple molecules, such as CO, H$_2$O and N$_2$ are reformed. On the other hand, some molecules are effectively formed at high temperature; carbon-chains, such as C$_2$H$_2$ and cyanopolyynes, are formed at the temperature of $>$700 K. We also find that large organic molecules, such as CH$_3$OH and HCOOCH$_3$, are associated with the first core ($r \lesssim 10$ AU). Although the abundances of these molecules in the first core stage are comparable or less than in the protostellar stage (hot corino), reflecting the lower luminosity of the central object, their column densities in our model are comparable to the observed values toward the prototypical hot corino, IRAS 16293-2422. We propose that these large organic molecules can be good tracers of the first cores.
\end{abstract}





\section{INTRODUCTION}
It is well established that a star is formed by gravitational collapse of a molecular cloud core. The collapsing core is initially optically thin to the thermal emission of dust grains, and undergo isothermal run-away collapse as long as the cooling rate overwhelms the compressional heating. The isothermal condition breaks down when the central density reaches $\sim$10$^{10}$ cm$^{-3}$, and the temperature starts rising. Increasing gas pressure decelerates the contraction, and the core come to the hydrostatic equilibrium, which is called a first hydrostatic core \citep[e.g.,][]{larson69,masunaga98}. Although the first core is a transient object, it plays essential roles in determining the physical condition of a protostar; it fragments and forms binaries \citep{matsumoto03}, and drives bipolar molecular outflows \citep{tomisaka02}. When the temperature reaches $\sim$2000 K and hydrogen molecules start to dissociate, the central region of the first core collapses again to form the protostar.

Star formation processes include both the structure evolution as mentioned above, and molecular evolution. Rates of chemical reactions depend on various physical quantities. Among them, the temperature is crucial; e.g., rates of ice sublimation and reactions with potential barriers are proportional to $\exp(-\Delta E/kT)$, where $\Delta E$ represents desorption energy or a height of an energy barrier. In addition, the molecular evolution in star forming regions is a non-equilibrium process. Therefore, accurate temperature determination via radiation hydrodynamic (RHD) simulations is important to investigate molecular evolution in star forming cores.

There are a few previous works which combined a chemical reaction network model with a RHD model of star-forming cores. Aikawa et al. (2008, here after AW08) investigated chemistry from a molecular cloud core to a protostellar core adopting a spherically symmetric RHD model \citep{masunaga00}, and mainly discussed the molecular evolution that occurs at $r \gtrsim 10$ AU, where $T \lesssim 300$ K. The spherical symmetry, however, should brake down inside the centrifugal radius of $\sim$100 AU, where circumstellar disks will appear. \citet{weeren09} adopted an axis symmetric RHD model \citep{yorke99}, and investigated molecular evolution mainly at $T < 100$ K from a collapsing molecular cloud cores to a forming circumstellar disk. The spatial resolution of their physical model is $\sim$3 AU in the central region. So the central first core and/or protostar is not resolved.
 
 
In this paper, we combine a detailed chemical reaction network model of gas-phase and grain-surface chemistry with a three dimensional RHD simulation of a star-forming core for the first time, and investigate molecular evolution at $T$ = 10--2000 K. So far, a few groups have succcessfully performed three-dimensional R(M)HD simulations of a low mass star-forming core up to the first core stage \citep{white06,commercon10,tomida10a, tomida10b}. On the other hand, evolution beyond first cores in three dimension is more challenging and investigated only very recently by \citet{bate10, bate11}, \citet{tomida12a}, and \citet{tomida12b}. So we focus on chemistry in the first core stage in the present work. Our motivations are twofold.

Firstly, first cores are good targets for Atacama Large Millimeter/submillimeter Array (ALMA). Although over 40 years have passed since \citet{larson69} predicted the presence of first cores, they have not yet been detected. Observation of first cores is challenging; they are buried in dense envelopes, and have very compact structure ($\sim$10 AU) and short lifetime (a few 1000 yr). Recently, some candidates have been reported: IRS2E in L1448 \citep{chen10}, Per-Bolo 58 \citep{enoch10} and L1541-mm \citep{pineda11}. Observational properties of these objects, low bolometric luminosity $L < 0.1$ $L_{\bigodot}$ and cold spectral energy distribution, are consistent with the theoretical predictions of first cores \citep[e.g.,][]{masunaga98,saigo11}. On the other hand, IRS2E has a high velocity outflow ($\sim$25 km/s), and Per-Bolo 58 is considerably luminous at 24 $\mu$m and associated with a well-collimated outflow \citep{dunham11}, which contradict the theoretical prediction that the outflow from first cores is slow ($\lesssim$5 km/s) and not well collimated \citep{machida08}. Confirmation and further investigations of the candidates require observation of molecular emission lines. Therefore, chemical models of first cores are highly desired.

Secondly, recent hydrodynamic simulations have shown that the outer region of first cores might directly evolve to circumstellar disks, while the central region collapses to form protostars \citep{saigo08,machida10,machida11,bate10,bate11}. In that case, the chemical composition of the first cores would be the initial composition of the circumstellar disks. The Molecular evolution during the formation and evolution of the circumstellar disks is still an open question. Although most chemical models of the circumstellar disks have used abundances of the molecular cloud cores as initial abundances, its validity is questioned \cite[][]{weeren09,visser11}. In other words, there remains a missing link between chemistry in the molecular cloud cores and the circumstellar disks. Our detailed three-dimensional chemical-hydrodynamical model from the molecular cloud cores to the first cores is the first step to fill the gap.

The rest of the paper is organized as follows. In Section 2, we briefly describe our RHD simulations and chemical reaction network model. In Section 3, we discuss trajectories of fluid parcels in our RHD simulations. We show molecular evolutions in the assorted fluid parcels, and the spatial distributions of molecular abundances in the first core stage. In Section 4, we discuss molecular column densities and uncertainties in our reaction network model. We compare our model with models of hot corinos in terms of physical and chemical properties. We also discuss the effect of accretion shocks on chemistry, and chemistry after the first core stage. We summarize our results in Section 5.

\section{MODEL}
\subsection{Physical Model: A 3 Dimensional Radiation Hydrodynamic Simulation}
Physical structure of a collapsing molecular cloud core is calculated using a three-dimensional nested-grid self-gravitational radiation magnetohydrodynamics code, which was developed by \citet{tomida10a}. Here we briefly describe physical parameters and structure of a core. 

The self-gravitational magnetohydrodynamics equations are coupled with radiation transfer, which is treated with gray flux-limited diffusion approximation \citep{levermore81}. The gas temperature is set to be equal to the dust temperature. In the present work, the magnetic field is set to zero for simplicity. Idealized equation of state is assumed with the adiabatic index of 5/3. The simulation consists of multiple nested-grids in order to achieve higher resolution in the central region.  Each level of the nested grids has 64$^3$ cubic cells in $(x, y, z)$. At the end of the simulation, 13 levels of nested-grids are generated and the smallest scale corresponds to $\sim$0.05 AU.

We calculate the evolution of a molecular cloud core with the mass of 1 $M_{\odot}$. The initial density distribution is given by a critical Bonnor-Ebert sphere, and enhanced by a factor of 1.6 so that the sphere becomes gravitationally unstable. Initially, the central number density of the core is $n_c \sim 8.3 \times 10^{5}$ cm$^{-3}$ ($\rho_c = 3.2 \times 10^{-18}$ g cm$^{-3}$). The radius of the cloud core is 6300 AU. The initial angular velocity is $2.0\times10^{-14}$ s$^{-1}$, and the initial temperature is set to 10 K. The simulation is stopped when the maximum temperature reaches 2000 K, at which temperature hydrogen molecules collisionally dissociate, and the second collapse starts.

\begin{figure}
\epsscale{0.8}
\plotone{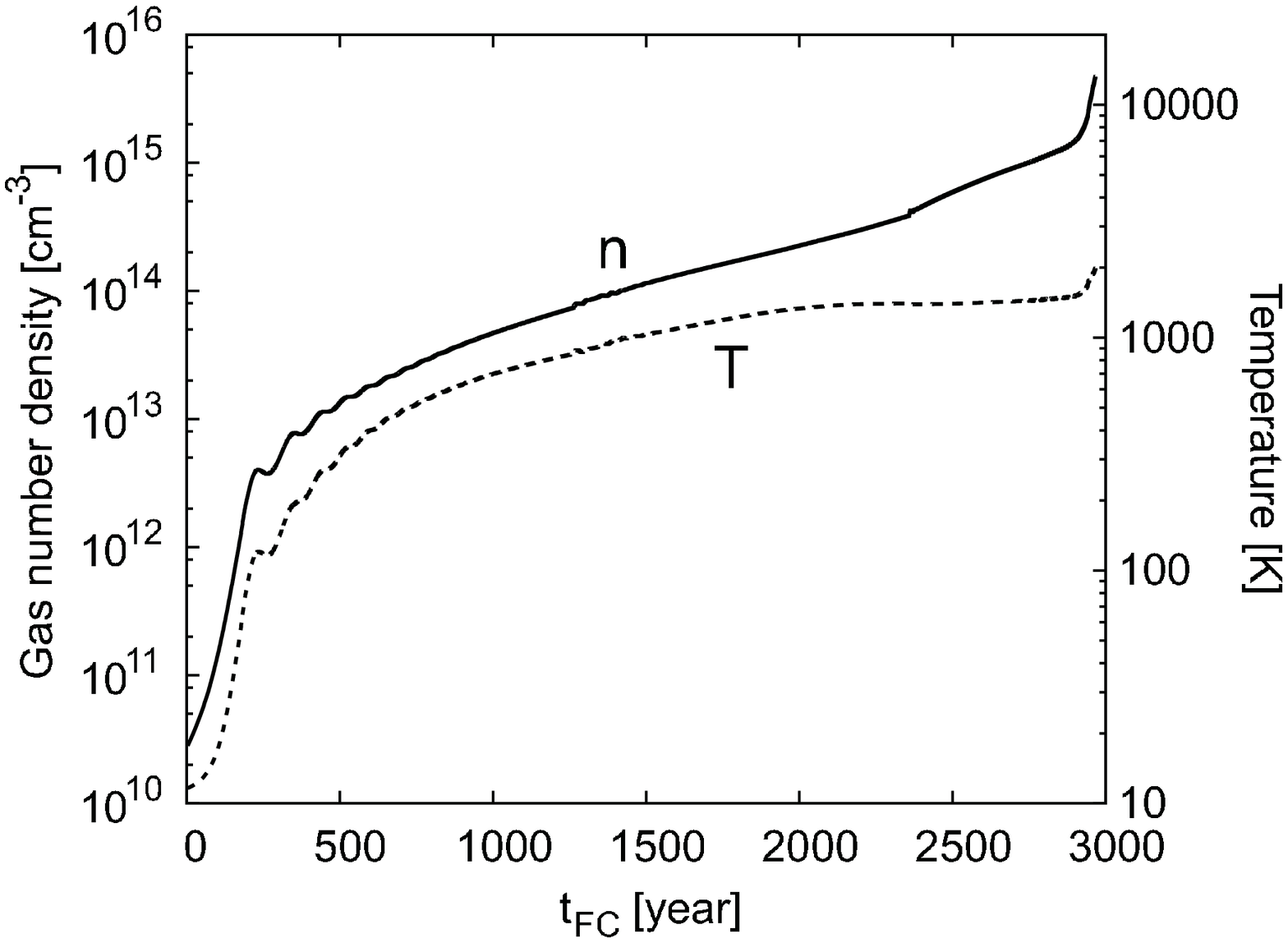}
\caption{Temporal variations of gas number density (solid line) and temperature (dashed line) at the center of the core.\label{fig1}}
\end{figure}

In this paper, we define the moment when the central gas number density of the core reaches $n_c \sim 2.6 \times 10^{10}$ cm$^{-3}$ ($\rho_c = 10^{-13}$ g cm$^{-3}$) as t$_{\rm FC}$ = 0; the first core is born. Figure \ref{fig1} shows the temporal variations of the central gas number density and temperature of the core at t$_{\rm FC} >$ 0. The temperature and the density increase by two and five orders of magnitude in $\sim$3000 yr, respectively, which is the lifetime of the first core in our model. Figure \ref{fig2} shows the core structures at $t_{\rm FC} = 1500$ yr (middle stage) and 2800 yr (late stage). We present molecular distributions in these two stages in Section 3. The core is oblate or disk-like because of the rotation. In the panels (c) and (f), the dashed lines depict temperature contours of 25 K and 100 K, which are roughly the sublimation temperature of CO and large organic molecules, respectively. As the core evolves from 1500 yr to 2800 yr, the sublimation radii of CO and large organic molecules expand from $\sim$40 AU to $\sim$100 AU and from $\sim$5 AU to $\sim$10 AU, respectively. We discuss the physical structures of the core in more detail in Section 3.1.

Chemical processes are not included in the hydrodynamic simulation. In this work we calculate molecular abundances as a post process; we trace 10$^5$ fluid parcels in the hydrodynamic simulation, and calculate a chemical reaction network model along 10$^4$ selected trajectories. Our approach is similar to AW08 and \citet{weeren09}. Initially, the fluid parcels are distributed spherically at $r =$ 1000--2500 AU. The parcels follow the local flow at each hydrodynamic time step. Leapfrog integration is used for tracing the parcels, and the physical parameters at the position of the parcels are obtained by the trilinear interpolation. In the calculations of molecular evolution in the fluid parcels, the temperature, density, visual extinction, and cosmic-ray ionization rate are updated at each time step along the trajectories (see also Section 2.2.4). We note that the accretion shock heating layer at the first core surface is not spatially resolved in our model, since the thickness of the layer, $\sim$10$^{-3}$ AU  \citep{saigo11}, is smaller than the smallest scale of our calculation. Effect of the shock layer on chemistry is discussed in Section 4.4.


\subsection{Chemical Reaction Network}
The dynamical evolution covers a wide range of temperature and density: 10--2000 K and 10$^5$--10$^{16}$ cm$^{-3}$ (see Figure \ref{fig1}). To calculate the molecular evolution in these physical conditions, we combine two chemical reaction network models, \citet{garrod06} and \citet{harada10}, as a base of our model. The former is a gas-grain reaction network model applied to a star-forming core at 10--200 K. The latter is a high temperature gas-phase reaction network model, and originally developed for the temperature of 100--800 K. The major difference between the two network models is that reactions with a high potential energy barrier are included in the latter model. For the neutral-neutral reactions and neutral-ion reactions, we switch the rates from the former network to the latter network when the temperature exceeds 100 K. In the following, we describe modifications to gas phase reactions, treatment of gas-grain interactions, and grain-surface reactions in our model. In total, the reaction network consists of 461 gas-phase species (including dust grains with three different charge states) and 195 grain-surface species, and 11696 reactions. The numerical code for solving rate equations is adapted from Nautilus \citep{hersant09,semenov10}.

\subsubsection{Modifications to Gas Phase Reactions}
Firstly, we modify branching ratios of dissociative recombination reactions. \citet{geppert06} suggested that formation of saturated complex molecules by dissociative recombination of their protonated precursors is not efficient. For example, when protonated methanol (CH$_3$OH$_2^+$) recombines, the branching ratio for CH$_3$OH + H is only 3\%. Recently, \citet{hamberg10} also showed that in dissociative recombination of dimethyl ether cation (CH$_3$OCH$_4^+$), the branching ratio for CH$_3$OCH$_3$ + H is only 7\%. Referring to these experiments, the network of \citet{garrod08}, which is an extension of \citet{garrod06} to more complex species, assumed that two-fragment channels (ex. X$^+$ + e$^-$ $\rightarrow$ Y + Z) represent 5\% each, and that the remainder is evenly split between channels with three or more fragments, while in the network of \citet{garrod06} the branching ratios of dissociative reactions of protonated complex molecules are equally weighted for most species. We modify the branching ratios in the recombination of protonated precursors of saturated complex molecules following \citet{garrod08}.

We add some three-body association reactions and collisional dissociation reactions for all neutral species to our network referring to \citet{willacy98} (see Appendix A). Collisional dissociation of H$_2$ is an exception; we use the rates in \citet{harada10} instead of \citet{willacy98}. Since collisional dissociation is endothermic, the rate of the collisional dissociation reactions is proportional to $\exp(-\gamma/T)$. The energy barrier, $\gamma$, must be larger than $\Delta H$, which is the difference between the formation enthalpy of products and reactants. In \citet{willacy98} the value of $\gamma$ is set to be lower than $\Delta H$ for some reactions. For those reactions, we calculate $\Delta H$ at 300 K referring to NIST-JANAF Thermochemical Tables (http://kinetics.nist.gov/janaf/) and KInetic Database for Astrochemistry (KIDA; http://kida.obs.u-bordeaux1.fr; Wakelam et al. 2012), and replace $\gamma$ with $\Delta H$. For the species which are not listed in the network of \citet{willacy98} the collisional dissociation rates are set to be, 
\begin{equation}
k_{cd} = 1.0\times 10^{-10}\exp(-\Delta H/T).
\end{equation}
If the formation enthalpy of reactants or products are not available in the literature, we assume $\Delta H$ is $5\times10^4$ K $(\sim$5 eV). We assume that the collisional dissociation reactions preserve functional groups. Otherwise, we assume that one H atom is extracted via the collisional dissociation. It should be noted that the set of collisional dissociation reactions in our model is likely incomplete because of the lack of experimental data; for many reactions listed in Table A2, there is no laboratory experiments in which products are confirmed. If the products are different from those listed in Table A2, the value of $\gamma$ should also be different. In Section 4.2, we discuss how the molecular abundances depend on the assumption on the collisional dissociations.

\subsubsection{Gas-Grain Interactions}
In molecular cloud cores, dust grains are negatively charged. But the population of charged particles changes as the gas density increases \citep{umebayashi83}. Grains become the dominant charged particles at high densities ($n \gtrsim 10^{10}$ cm$^{-3}$), and eventually grain-surface recombination (ion-grain collision) overwhelms gas-phase recombination (recombination of an free electron with an ion). Grain charges should therefore be calculated in the network simultaneously.
 
We consider neutral and singly charged dust grains. Gaseous species and dust grains interact through electron-grain, ion-grain collisions and adsorption and desorption of neutral species. Table 1 lists the generalized scheme of gas-grain interactions in our model.

A collision between a negatively-charged grain and a positively-charged grain result in two neutral grains. The collision rate of a grain of radius $a$ and charge $le$ with another grain of radius $a'$ and charge $l'e$ is given by
\begin{equation}
k_{gg} = \pi(a+a')^2\left(\frac{8kT}{\pi\mu_g}\right)^{1/2}\left[1-\frac{ll'e^2}{(a+a')kT}\right],
\end{equation}
where $\mu_g$ is the reduced mass of the colliding grains \citep{umebayashi83}. We simply assume all grains have the same radius of 10$^{-5}$ cm and density of 3 g cm$^{-3}$. In the case of electron-grain collisions, the rates are given by
\begin{equation}
k_{eg} = S_e\pi a^2 \left(\frac{8kT}{\pi m_e} \right)^{1/2}\tilde{J} \left(\tau = \frac{akT}{e^2}, \nu = -l \right),
\end{equation}
where $m_e$ is the mass of electron. An electron sticks to a grain upon collision with a sticking probability of $S_e$, which strongly depends on the temperature. We calculate $S_e$ following \citet{umebayashi83}, assuming a potential depth of the grain equals to 2 eV. $\tilde{J}$ is a dimensionless reduced rate coefficient including the polarization effect of the grains by the electric field of approaching species. We applied the approximate formulas given by \citet{draine87},
\begin{equation}
\tilde{J}(\tau, \nu=0) = 1+ \left(\frac{\pi}{2\tau}\right)^{1/2},
\end{equation}
\begin{equation}
\tilde{J}(\tau, \nu<0) = \left[1-\frac{\nu}{\tau}\right] \left[1+\left(\frac{2}{\tau-2\nu}\right)^{1/2}\right].
\end{equation}




The collision of a cation with a negatively charged grain is theoretically discussed in \citet{watson72}, and summarized by \citet{umebayashi80}. According to these references, there are two possible processes when a cation approaches a negatively charged grain; a free electron tunnels to the cation before it hits the grain, or the recombination occurs on the grain surface. For simplicity, we assume the former case, in which the products and branching ratios of the recombination would be the same as the recombination in the gas phase. In the case of the collision of a cation with a neutral grain, recombination probably occurs on grain surfaces, and the products desorb to the gas phase, using the excess energy (T. Umebayashi 2012, private communication). Again, we assume that the products and branching ratios are the same as the corresponding recombination in the gas phase, for simplicity, although they might be different. For example, if X$^+$ + e$^-$ $\rightarrow$ Y + Z in gas-phase recombination, corresponding grain-surface recombinations are
\begin{eqnarray}
{\rm X^+ + gr^- \rightarrow Y + Z + gr},\\
{\rm X^+ + gr \rightarrow Y + Z + gr^+}.
\end{eqnarray}
Our network includes recombination in the gas phase and on grain surfaces for all cations. Since we assume that each branching ratio for two fragment channels is only 5\% (Section 2.2.1), our approach tends to produce smaller molecules via the dissociation of large species. If we assume that, for example, the branching ratios for all product channels are equal or the collisions lead to simple charge exchange (ex. X$^+$ + gr$^-$ $\rightarrow$ X + gr), formation rates of large organic species are enhanced. In other words, our model gives the lower limit of the abundances of large organic species. We discuss how much the abundances of large organic molecules are enhanced if we assume the equally weighted branching ratios for grain surface recombination in Section 4.2.

Recently, anions have been detected toward both the prestellar core, TMC-1 \citep{brunken07}, and the protostellar core, L1527 \citep{sakai07,agundez08}. Our network includes H$^-$, C$^-$, O$^-$, S$^-$, CN$^-$, and OH$^-$. Since interaction between anions and grains are unclear, we assume simple charge exchange,
\begin{equation}
{\rm X^- + gr^+ \rightarrow X + gr}.
\end{equation}
The rates of grain-ion (both cation and anion) collisions are expressed as 
\begin{equation}
k_{jg} = \pi a^2 \left(\frac{8kT}{\pi m_j} \right)^{1/2}\tilde{J} \left(\tau = \frac{akT}{q_j^2}, \nu = \frac{le}{q_j} \right),
\end{equation}
where $m_j$ and $q_j$ are the mass and the charge of ions, respectively.

Gaseous neutral species collide with dust grains and stick to the grain surface. AW08 and \citet{weeren09} assumed sticking probabilities of 0.5 and unity, respectively, independent of the temperature. Since the sticking probability should be lower at higher temperatures, we assume that the sticking probability for neutral species is
\begin{equation}
S_n = [1+0.04(T+T_{\rm d})^{0.5}+2\times 10^{-3}T + 8\times 10^{-6}T^{2}]^{-1}, \label{eqn:S_n}
\end{equation}
where $T$ and $T_{\rm d}$ are the gas and dust temperatures, respectively. We assume that $T$ and $T_{\rm d}$ are the same. Although Equation (\ref{eqn:S_n}) is originally derived for the sticking probability of a H atom \citep{hollenbach79} by fitting the numerical calculations, we use it for all neutral species. For example, the value of $S_n$ is 0.83 at 10 K and 0.54 at 100 K. We might underestimate $S_n$, because a loss of kinetic energy would be larger for collisions of heavier species with grains.

After species are adsorbed onto grain surfaces, they can desorb to the gas phase again. We adopt the same desorption energies as \citet{garrod06}, in which grain-surfaces are assumed to be covered with H$_2$O ice. In addition to the thermal desorption, we consider two non-thermal desorption processes. One is the cosmic-ray induced desorption; grains are heated up to 70 K by the cosmic-ray bombardment, and it allows temporal evaporations of species with low binding energies \citep{hasegawa93}. The other is the chemical desorption; energy released by the exothermic grain-surface reactions is used to desorb the products. The efficiency of the desorption is parameterized by $a_{\rm RKK}$, which is the ratio between the surface-molecule bond frequency and the frequency at which the energy is lost to the surface. We assume $a_{\rm RKK}=$ 0.01 referring to \citet{garrod07}; roughly 1\% of the products are desorbed into the gas phase.

\subsubsection{Grain Surface Reactions}
Calculation of grain-surface chemistry is performed in a similar way to \citet{garrod06}. The diffusion energy barrier is set to be a half of the desorption energy of each species. The diffusion rate is exponentially dependent on the diffusion barrier divided by the grain temperature, and surface reaction rates are determined by the sum of the diffusion rates for two reaction partners \citep{hasegawa92}. If surface reactions have activation energy barriers, the barriers are overcome thermally, or via quantum tunneling, whichever is faster. The modified rate method \citep{caselli98} is used only for reactions involving atomic hydrogen. Our model also includes photodissociation on grain surfaces by interstellar and cosmic-ray induced UV photons.


Our model assumes that molecules are physorbed, which is appropriate as long as grain surfaces are covered with ice. At $T \gtrsim 150$ K, however, ice mantle is sublimated, and chemisorption might become important \citep[e.g., Section 4.2.2 of][]{tielens05}. Since grain-surface reactions via chemisorption is out of the scope of this paper, we switch off adsorption of neutral gas-phase species onto grain surfaces, desorption of ice-mantle species to the gas phase, and grain surface reactions, if the temperature becomes greater than 200 K.


\subsubsection{Initial Molecular Abundances and Ionization Rates}
Table 2 lists the initial abundances in our model. The elemental abundances adopted here are called "low metal" values because of their strong depletions of species heavier than oxygen \citep[e.g.,][]{graedel82}. Species are initially assumed to be atoms or atomic ions except for hydrogen, which is in molecular form. We integrate the initial abundances for 10$^5$ yr with the typical conditions of dense molecular cores, $T = 10$ K, $n_{\rm H} = 1\times10^4$ cm$^{-3}$ and visual extinction of A$_V$ = 10 mag, which determine the abundances of our initial Bonnor-Ebert like sphere. Before the collapse begins, we further integrate the updated abundances under the initial hydrostatic conditions during $1.6\times10^5$ yr (i.e., sound crossing time of our core model), in order to obtain the starting abundances for the collapse, implicitly assuming that the core keeps its initial hydrostatic structure supported by turbulence. Our approach is similar to AW08 and van Weeren et al. (2009).

In the hydrodynamic simulation, we calculate column densities of gas in the $x-$, $y-$ and $z-$directions, $\Sigma_x$, $\Sigma_y$ and $\Sigma_z$, from the core outer edge to a position of each fluid parcel. We choose the minimum of $\Sigma_x$, $\Sigma_y$, $\Sigma_z$ and convert it to A$_V$, via the formula A$_V=(f_{\rm H}\Sigma/\mu m_{\rm H})\times(5.34 \times 10^{-22}$ cm$^{2}$) mag \citep{bohlin78,cardelli89}, where $f_{\rm H}$ is the mean number of hydrogen nuclei per species (1.67), $\mu$ is the mean molecular weight (2.3) and $m_{\rm H}$ is the mass of a hydrogen atom. For the fluid parcels we follow, the initial values of A$_V$ are greater than 10 mag, and A$_V$ increases as the collapse proceeds. Hence the interstellar UV radiation has little effect on chemistry in our model.

Cosmic rays are the main source of ionization in molecular clouds. The cosmic-ray ionization rate of hydrogen is set to be $1.3\times10^{-17}$ s$^{-1}$. The attenuation length of cosmic-ray is 96 g cm$^{-2}$ \citep{umebayashi81}, which is much larger than the column density of typical molecular clouds, but can be smaller than the column density, $\Sigma$, calculated above. The decay of short-lived radio active isotopes, which releases energetic particles, also ionize the ISM with rates $\xi$ = (7.6--11) $\times$ 10$^{-19}$ s$^{-1}$ \citep{umebayashi09}. Thus for direct cosmic-ray ionization and cosmic-ray-induced photo reactions, we assume that the effective cosmic-ray ionization rate is expressed as
\begin{equation}
\xi_{eff} = 1.3 \times 10^{-17}\exp \left(-\frac{\Sigma}{96\hspace{2pt}{\rm g\hspace{2pt}cm}^{-2}}\right) + 1 \times 10^{-18}.
\end{equation}
In our model, the decay of radio active isotopes is the main source of ionization within the first core, while the cosmic-ray dominates at the surface of the first core and infalling envelopes.  
\section{RESULT}
\subsection{Physical Structure of the Core}
Figure \ref{fig3} shows the spatial distributions of the gas number density, radial velocity ($v_r$), and temperature perpendicular to (a--c) and along the rotational axis (d--f). We averaged the distributions perpendicular to the rotational axis in the $z = 0$ plane in the azimuth direction. The solid lines represent the distributions at $t_{\rm FC} = 1500$ yr, while the dashed lines represent the distributions at $t_{\rm FC} = 2800$ yr. We define $r$ and $R$ as $\sqrt{x^2+y^2+z^2}$ and $\sqrt{x^2+y^2}$, respectively. Although the core has the rotation, the core structures are almost spherical at $r \gtrsim$ 10 AU both in the middle and late stages. The density has a power law distribution of $r^{-2}$ at $r \gtrsim 100$ AU and $r^{-1.5}$ at 10 AU $\lesssim r \lesssim 100$ AU. The effect of the rotation becomes apparent in the inner region ($r \lesssim 10$ AU) especially in the late stage, in which infalling gases have larger angular momentum. The accretion onto the first core is supersonic, and the accretion shock occurs at the first core surface. The first core is quasi-hydrostatic, and in the middle stage its radius is $\sim$6 AU perpendicular to the rotational axis and $\sim$5 AU along the rotational axis, respectively, while it is $\sim$10 AU and $\sim$3 AU in the late stage.

The temperatures both in the envelope and the first core increase with time as the first core grows, except for the low-temperature region in 5 AU $\lesssim R \lesssim$ 10 AU around the equatorial plane. This region experiences strong expansion perpendicular to the rotational axis in a dynamical timescale due to the angular momentum transport. Cooling by the expansion is more efficient than the radiation heating, since the gas is optically thick.

\subsection{Spatial Distribution of Fluid Parcels}
Figure \ref{fig4} shows the spatial distributions of fluid parcels in the middle (left) and late (right) stages. Each parcel is color-coded by its initial distance from the core center (above the equatorial plane) or its initial specific angular momentum (below the equatorial plane). In the middle stage, the first core is composed mainly of the fluid parcels which were initially at $r \sim$ 1150--1450 AU. In the late stage, these parcels have moved to the midplane of inner radii due to contraction, and the outer layers of the first core is composed of parcels falling from the initial radii of $r \sim$ 1700--1850 AU. It indicates that the accretion of the envelope onto the first core is active, and the gas inside the first core is continuously compress into the center. Parcels closer to the rotational axis have smaller initial angular momentum. While parcels fall almost radially in the envelope, they move in the $z$ direction in the first core rather than radially until they reach the equatorial plane. Near the equatorial plane, parcels move inward in the direction perpendicular to the rotational axis by losing the angular momentum.




\subsection{Molecular Evolution in a Single Fluid Parcel}
Here, we show the thermal and chemical histories of a single fluid parcel, which are helpful to understand the spatial distributions of molecules presented in Section 3.4. Figure \ref{fig5} shows the trajectory of the parcel, which is initially at $r = 1300$ AU  and 70$^\circ$ degrees from the rotational axis. The first core surface (white dashed line) is defined as the isoline of the minimum gas density in the region where the local sound speed is greater than the radial velocity. Figure \ref{fig6}(a) shows the temporal variations of the position of the parcel in $r, R, z$. The horizontal axes is set to be $t_{\rm final} - t$, where $t_{\rm final}$ represents the final stage of our model; $t_{\rm FC}$ = 0 corresponds $t_{\rm final} - t \sim 3000$ yr. The vertical dotted lines in Figure \ref{fig6} represent the time when the parcel passes the accretion shock at the first core surface ($t_{\rm final} - t = 1930$ yr). The parcel initially infalls almost spherically. The effect of the rotation becomes apparent inside the first core, and the trajectory becomes helical. The parcel falls to the equatorial plane within 10$^2$ yrs after entering the first core, while it moves inward in the direction perpendicular to the rotational axis over $10^3$ yrs by losing its angular momentum. 

Figure \ref{fig6}(b) shows the temporal variations of the temperature in the parcel. We classify the history of the parcel into three phases in terms of the temperature; cold phase (10--20 K), warm-up phase (20--200 K) and hot phase ($>$200 K). Dominant chemical processes vary among the three phases. In the cold phase, adsorption onto and hydrogenation on grain surfaces are dominant, while in the warm-up phase, the molecular evolution is caused by grain-surface reactions among species with heavy elements and evaporation of ice mantle species. In the hot phase, the molecular evolution is caused by gas phase reactions. This classification has been used in the chemistry in protostellar stage \citep[e.g.,][]{herbst09}. It should be noted, however, that in the first core stage the density is higher and the duration of the warm-up and hot phases are shorter than in protostellar stage. Table 3 summarizes the physical conditions and duration of the three phases in the parcel.

In the following, we show the molecular evolution in these three phases. However, the temporal variations at 20 K $\lesssim T \lesssim$ 1000 K is difficult to see because the temperature rises in a short time scale ($\sim$10$^3$ yrs). In order to zoom-up the evolution at that temperature range, each panel in Figures \ref{fig6}(c--h) is divided into three phases based on the above classification. The panels (c), (d) and (e--h) show the number density of H nuclei, temperature, and molecular abundances relative to H nuclei in the parcel, respectively. The physical conditions as well as the molecular abundances look discontinuous at the boundaries of the phases, which is due to the different scaling of the horizontal axis. We assume that the parcels stay for $1.6\times10^5$ yr in its initial position, as mentioned in Section 2.2.4, but this pre-collapse phase is omitted in Figure \ref{fig6}.

\subsubsection{Cold Phase}
When the core begins to contract ($t_{\rm final}-t = 7\times10^4$ yr), molecules with heavy elements are mostly exists on grain surfaces. The main carbon reservoirs are CO, CH$_4$ and H$_2$CO in order of high to low abundances. Water, which is formed by hydrogenation of oxygen atoms on grain surfaces, is as abundant as $\sim$10$^{-4}$. The nitrogen mainly exists as N$_2$ and NH$_3$. All these molecules mostly exist on grain surfaces.

At the beginning of the contraction, the parcel migrates inward almost spherically and isothermally. As the parcel falls to inner denser regions, the abundances of gaseous molecules decrease due to freeze-out. Hydrogenation on grain surfaces occur subsequently, and the abundances of formaldehyde (H$_2$CO) and methanol (CH$_3$OH) slightly increase. At the end of the cold phase, the abundances of CH$_3$OH and H$_2$CO are $3\times10^{-6}$ and $1\times10^{-5}$, respectively.

\subsubsection{Warm-up Phase}
After the first core is formed, the parcel continues to infall, and the temperature in the parcel starts to rise. The parcel passes the accretion shock at the first core surface at $t_{\rm final}-t = 1930$ yr, when the number density of H nuclei is $3\times10^{10}$ cm$^{-3}$ and the temperature is 80 K (vertical dotted lines in Figure \ref{fig6}). Since the shock heating layer is not spatially resolved in our hydrodynamic simulation, our chemical model neglect this shock heating. We discuss the effect of the accretion shock on chemistry in Section 4.4.

During the warm-up phase, species are desorbed into the gas phase when the temperature reaches their sublimation temperatures. CO is desorbed to the gas phase when the temperature reaches 25 K, which is slightly higher than the typical interstellar value of 20 K, because of the higher density ($\sim$10$^{9}$ cm$^{-3}$) of the envelope in the first core stage. On the other hand, species with the higher binding energies are not desorbed yet; they migrate on grain surfaces and react with each other at the lukewarm temperature of $T \sim$ several 10 K. AW08 found that large organic molecules, such as HCOOCH$_3$ and CH$_3$OCH$_3$ are efficiently formed at $T =$ 20--40 K, and reach the terminal abundances of $\sim$10$^{-9}$ in the protostellar core. In our model, HCOOCH$_3$ is formed at $T =$ 30--50 K, but reaches abundances of only $2\times10^{-11}$. The abundances of formic acid ($6\times10^{-11}$) and CH$_3$CN ($6\times10^{-9}$) change little in the warm-up phase. The formation of large organic species in the warm-up phase is inefficient in our model because of the short time scale spent in the lukewarm region, which is determined by the size of the lukewarm region divided by the infalling velocity: it is only $\sim$10$^2$ yrs in our model, while it continues $\sim$10$^4$ yrs in AW08 \citep[see also][]{garrod06}. These large organic molecules start to be desorbed at $T > 100$ K.

\subsubsection{Hot Phase}
In the hot phase, ice mantles of grains are completely evaporated, and gas phase reactions determine the molecular abundances. The temperature and density of H nuclei reach 2000 K and $8\times10^{15}$ cm$^{-3}$, respectively at the final step. The duration of the hot phase is $\sim$$2\times10^3$ yrs.


At the temperature of 200--500 K, the composition stays almost the same as in the warm-up phase. At $T \gtrsim 500$ K, most molecules are destroyed, and simple molecules are reformed. For example, H$_2$CO starts to be destroyed by collisional dissociation when the temperature reaches $\sim$500 K. CH$_3$OH is also collisionally dissociate at $T \sim 1200$ K. Methane, which is the second carbon reservoir, is converted to CO by neutral-neutral reactions in several hundred yrs. As a result, the abundance of CO increases by a factor of 2. The abundance of H$_2$O, which is the main oxygen reservoir, remains the almost constant value of $\sim$10$^{-4}$, which is mostly determined in the precollapse phase. The dominant reservoirs of the nitrogen are N$_2$ and NH$_3$. At $T \gtrsim$ 1600 K, NH$_3$ is converted to N$_2$ through the reactions of NH$_3$ + H $\rightarrow$ NH$_2$ + H$_2$, and NH$_2$ + NO $\rightarrow$ N$_2$ + H$_2$O on the time scale of $\sim$10 yrs. The former reaction has the potential energy barrier of $5.2\times10^3$ K. The abundance of hydrogen atom increases with time mainly by collisional dissociation of H$_2$. When the temperature reaches $\sim$2000 K, $\sim$0.1\% of hydrogen nuclei is in hydrogen atom. As the abundance of hydrogen atom increases, OH is formed through the reaction, H$_2$O + H $\rightarrow$ OH + H$_2$, which has the potential energy barrier of $9.7\times10^3$ K.

Some larger molecules are temporary formed in the gas phase. Methyl formate and CH$_3$OCH$_3$ are formed by the grain-surface recombination of HCOOCH$_4^+$ and CH$_3$OCH$_4^+$, respectively, despite the small branching ratio of 5\%. These formation paths are efficient until CH$_3$OH, which is the precursor of HCOOCH$_4^+$ and CH$_3$OCH$_4^+$, is destroyed by collisional dissociation. The peak abundances of HCOOCH$_3$ and CH$_3$OCH$_3$ are $5\times10^{-11}$ and $9\times10^{-11}$, respectively.
At the temperature of $\sim$700 K, acetylene (C$_2$H$_2$) is formed by the reaction of H$_2$ and C$_3$H$_4$, which is formed on grain surfaces during the cold phase. The reaction has the potential barrier of $2.3\times10^4$ K. The abundance of C$_2$H$_2$ reaches $\sim$10$^{-6}$ at the maximum. C$_2$H$_2$ is destroyed through the reactions of C$_2$H$_2$ + OH $\rightarrow$ CH$_2$CO + H, and CH$_2$CO + M $\rightarrow$ CH$_2$ + CO, where M represents the third body (mainly H$_2$). These reactions have the potential energy barrier of $5.0\times10^3$ K and $3.9\times10^4$ K. Other unsaturated carbon-chain molecules, C$_x$H$_y$, are temporally formed via the neutral-neutral reactions, C$_x$H$_{y+1}$ + H $\rightarrow$ C$_x$H$_{y}$ + H$_2$, which have the potential energy barrier of several 10$^3$ K. HCN is converted to HC$_3$N through the reactions of HCN + H $\rightarrow$ CN + H$_2$, and C$_2$H$_2$ + CN $\rightarrow$ HC$_3$N + H. The former reaction has the potential energy barrier of $1.2\times10^4$ K. Larger cyanopolyynes (HC$_n$N, $n = 5, 7, 9$) are also formed by the subsequent neutral-neutral reactions of HC$_3$N + C$_2$H, HC$_5$N + C$_2$H and HC$_7$N + C$_2$H. The terminal abundances of HC$_n$N ($n = 3, 5, 7, 9$) are $4\times10^{-7}$, $1\times10^{-7}$, $2\times10^{-8}$ and $4\times10^{-9}$, respectively. These cyanopolyynes survive until the end of the simulation, because the neutral-neutral reactions, ${\rm C}_n{\rm N} + {\rm H}_2 \rightarrow {\rm HC}_n{\rm N} + {\rm H}$, which have the potential barriers of 2000 K, are compensated by the formation via the collisional dissociation reactions, ${\rm HC}_n{\rm N} + {\rm M} \rightarrow {\rm C}_n{\rm N} + {\rm H} + {\rm M}$. Therefore, the abundances of cyanopolyynes depend on the products of the collisional dissociation reactions assumed in our model.

In summary, the total molecular abundances (gas and ice combined) of many species in the cold phase mostly remain unaltered until the temperature reaches 500 K. Above 500 K, the molecules with high abundances start to be destroyed, and are reformed to simple molecules, such as CO, H$_2$O and N$_2$. Some molecules are formed in the warm-up phase and the hot phase; HCOOCH$_3$ is formed on grain surfaces during the warm-up phase, while it is formed in the gas phase during the hot phase. Carbon-chains, such as C$_2$H$_2$ and cyanopolyynes, are effectively formed in the gas phase in the hot phase.

\subsection{Spatial Distribution of Molecular Abundances}
In order to derive the spatial distributions of molecular abundances, we calculate molecular evolution in 10$^4$ parcels, which are selected from the 10$^5$ parcels traced in our RHD simulations. Figure \ref{fig7} shows the distributions of the parcels, which are color-coded according to the fractional abundances of gaseous CH$_3$OH. The contour lines and gray scales depict the temperature and number density, respectively, in logarithm scale. The abundance of CH$_3$OH jumps at $r \sim 5$ AU in the middle stage and $r \lesssim 10$ AU in the late stage, where the temperature reaches its sublimation temperature ($\sim$100 K). At the outer radii, CH$_3$OH exists on grain surfaces. The spatial distributions of other many molecules are similar, while sublimation radii vary among species. In the late stage, the abundance of CH$_3$OH drops in the central region ($r \lesssim$ 1 AU), because CH$_3$OH is collisionally dissociated.

Figure \ref{fig7} indicates that the distributions of CH$_3$OH abundance follows the local temperature. To confirm this, we plot the abundance of CH$_3$OH as a function of temperature in the middle stage (gray square) and the late stage (black cross) in Figure \ref{fig8}. It clearly shows that the CH$_3$OH abundances follow the local temperatures. Note, however, the plots in the middle stage and the late stage are not identical. It indicates that the fluid parcels with the similar temperature $at$ $each$ $evolutionary$ $stage$ have experienced similar thermal histories and molecular evolutions. At the temperature of around 100 K and 1200 K, the CH$_3$OH abundances are scattered. The CH$_3$OH abundances are determined by the sublimation at $\sim$100 K and collisional dissociation at $\sim$1200 K. The rates of these processes have the exponential dependence on the temperature, and a little difference of the thermal histories may make the scattering. In the late stage, there are two branches at the temperature below $\sim$100 K. The lower branch represents the CH$_3$OH abundances in the local low-temperature region inside the first core, which is mentioned in Section 3.1, while the upper branch represents the abundances in the envelope. Since the density is much higher in the first core than in the envelope, the degree of the depletion is much sever in the first core.

The spatial positions of the parcels at each snapshot constitute an irregular grid for which the molecular abundances, the density and the temperature are known. We have interpolated the molecular abundances in the density-temperature space with the nearest neighbors method, and obtained the table of the molecular abundances as the function of the density and the temperature at each snap shot. These tables are used to derive the molecular abundances at each evolutionary stage at the points of three-dimensional regular Cartesian-grids for which the temperature and the density are known from our hydrodynamic simulation. We use these data to derive the profiles of molecular abundances and column densities. Since the molecular evolution in the star-forming regions is a non-equilibrium process, in general, the abundances are not given as a function of only the local physical conditions, but also depend on the past physical conditions. But in our case, the abundances can be given as a function of the local temperature and density $at$ $each$ $evolutionary$ $stage$, because the flow pattern in our hydrodynamic simulation is simple, and again the fluid parcels with the similar physical conditions at each evolutionary stage have experienced the similar thermal histories and molecular evolutions.

\begin{figure}
\epsscale{0.6}
\plotone{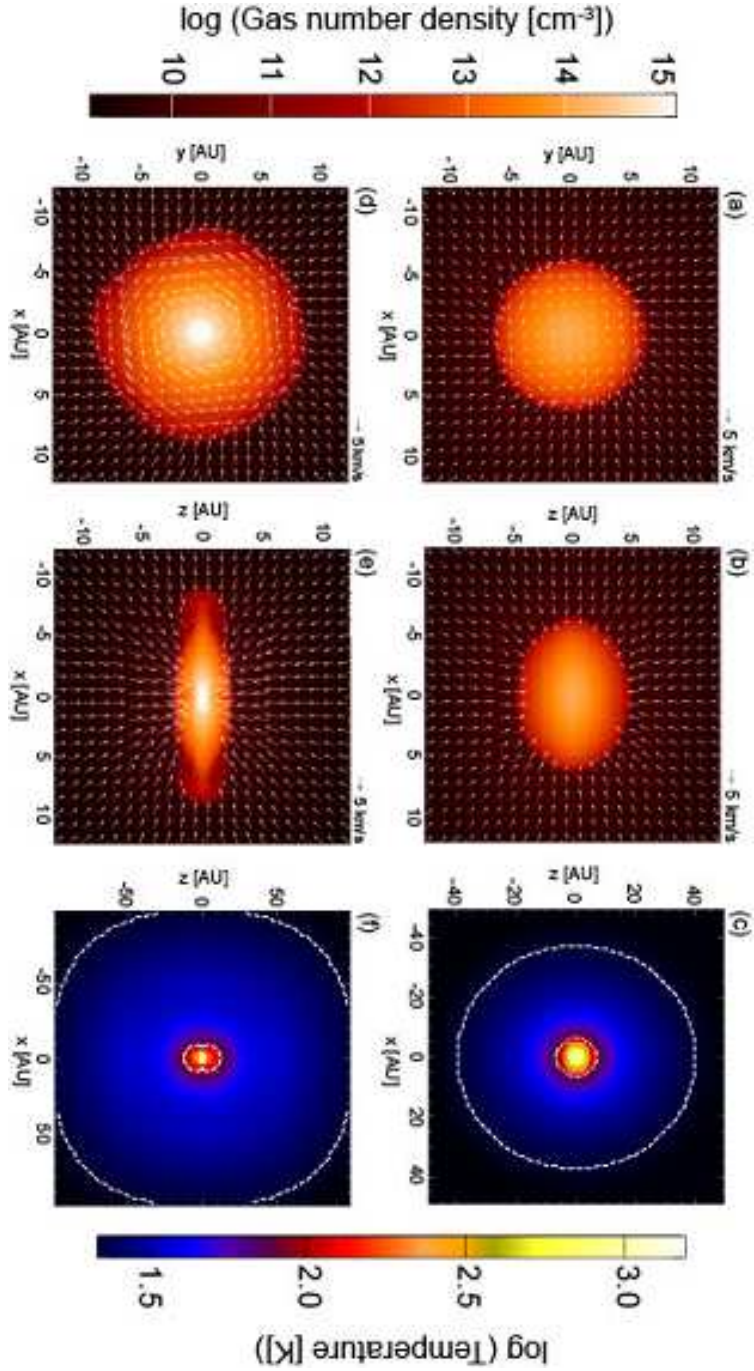}
\caption{Distributions of gas number density (a, b, d, e) and temperature (c, f) at $t_{\rm FC} = 1500$ yr (a--c) and 2800 yr (d--f). The panels (a) and (d) represent the cross sections at $z = 0$, while the other panels represent the cross sections at $y = 0$. The arrows denote the velocity fields. The white dashed lines in the panels (c) and (f) depict isothermal lines of 25 K and 100 K.
 \label{fig2}}
\end{figure}

\begin{figure}
\epsscale{0.6}
\plotone{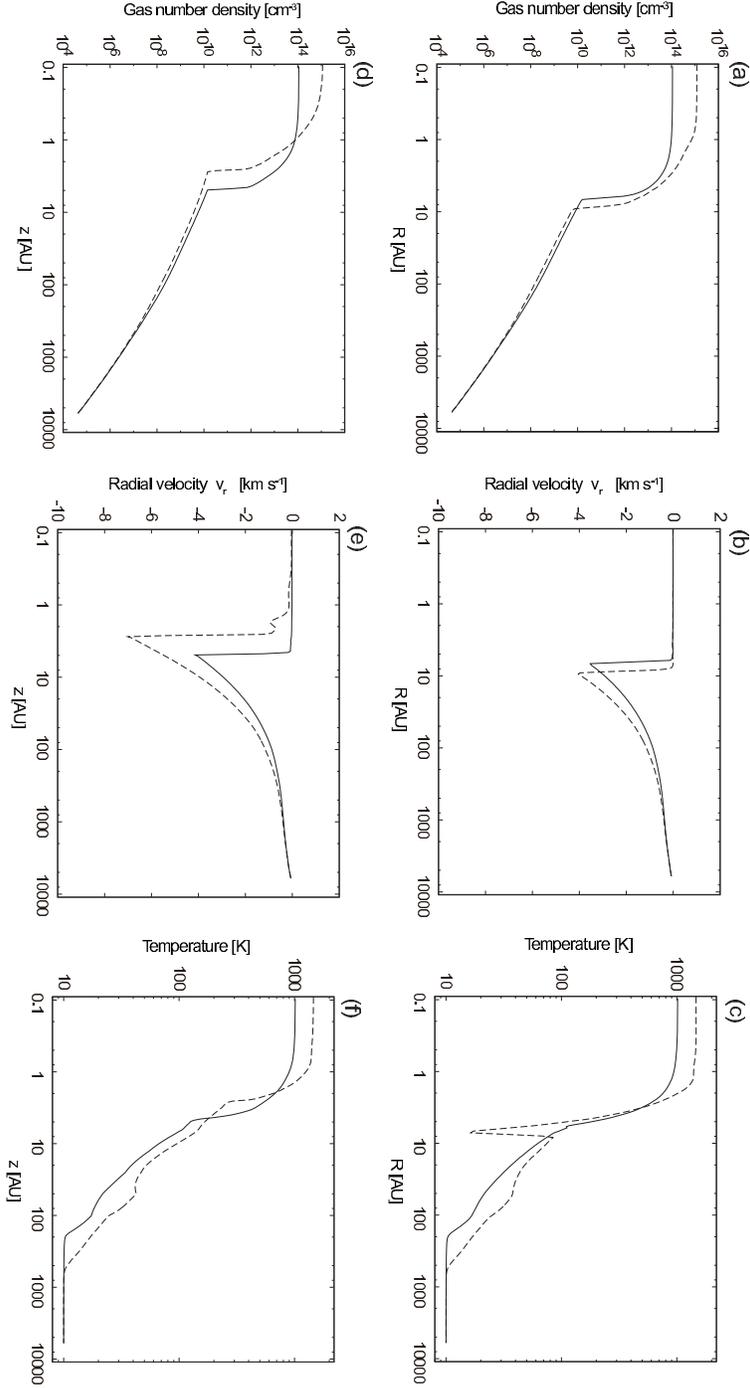}
\caption{Distributions of gas number density, radial velocity ($v_r$) and temperature perpendicular to (a--c) and along the rotational axis (d--f) at $t_{\rm FC} = 1500$ yr (solid lines) and 2800 yr (dashed lines). Distributions perpendicular to the rotational axis are in the $z = 0$ plane and averaged in the azimuth direction. \label{fig3}}
\end{figure}

\begin{figure}
\epsscale{0.48}
\rotatebox{90}{\plotone{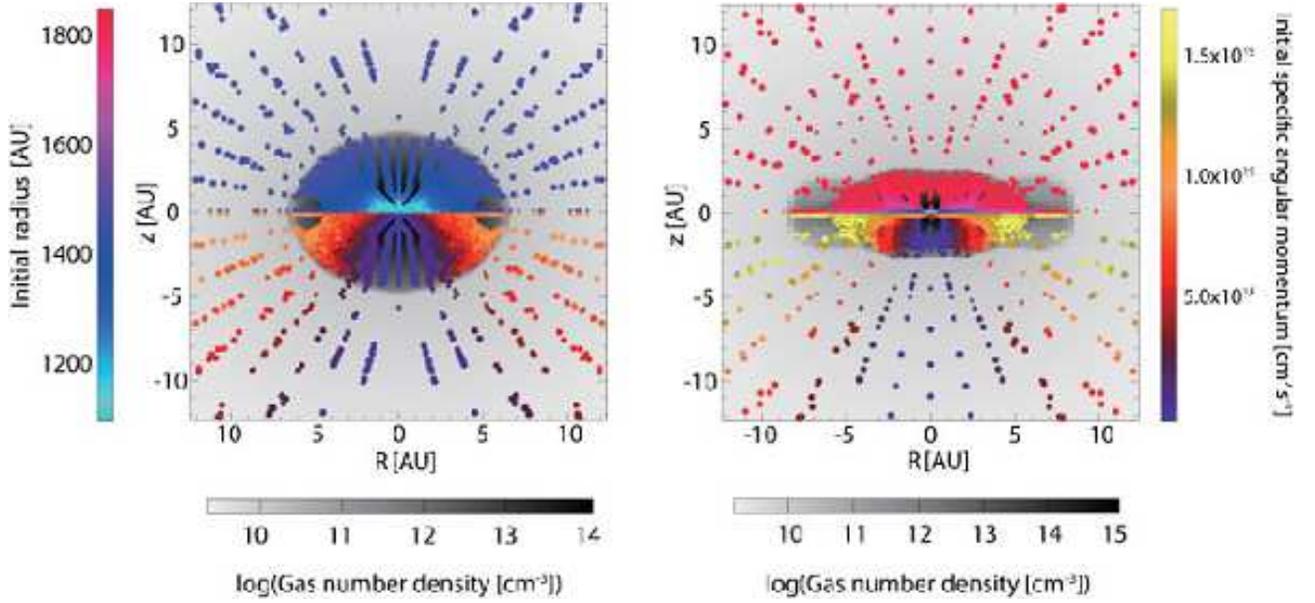}}
\caption{Spatial distributions of fluid parcels in the middle stage (left) and the late stage (right). The parcels are color-coded according to their initial radius (above the equatorial plane) and initial specific angular momentum (below the equatorial plane). Gray scale shows the distributions of gas number density averaged in the azimuth direction. A parcel is plotted at $(R, z) = (\sqrt{x_p^2 + y_p^2}, z_p)$, when the position of the parcel is $(x_p, y_p, z_p)$. $R < 0$ regions are simply copied from $R > 0$ regions.\label{fig4}}
\end{figure}

\begin{figure}
\epsscale{1.0}
{\plotone{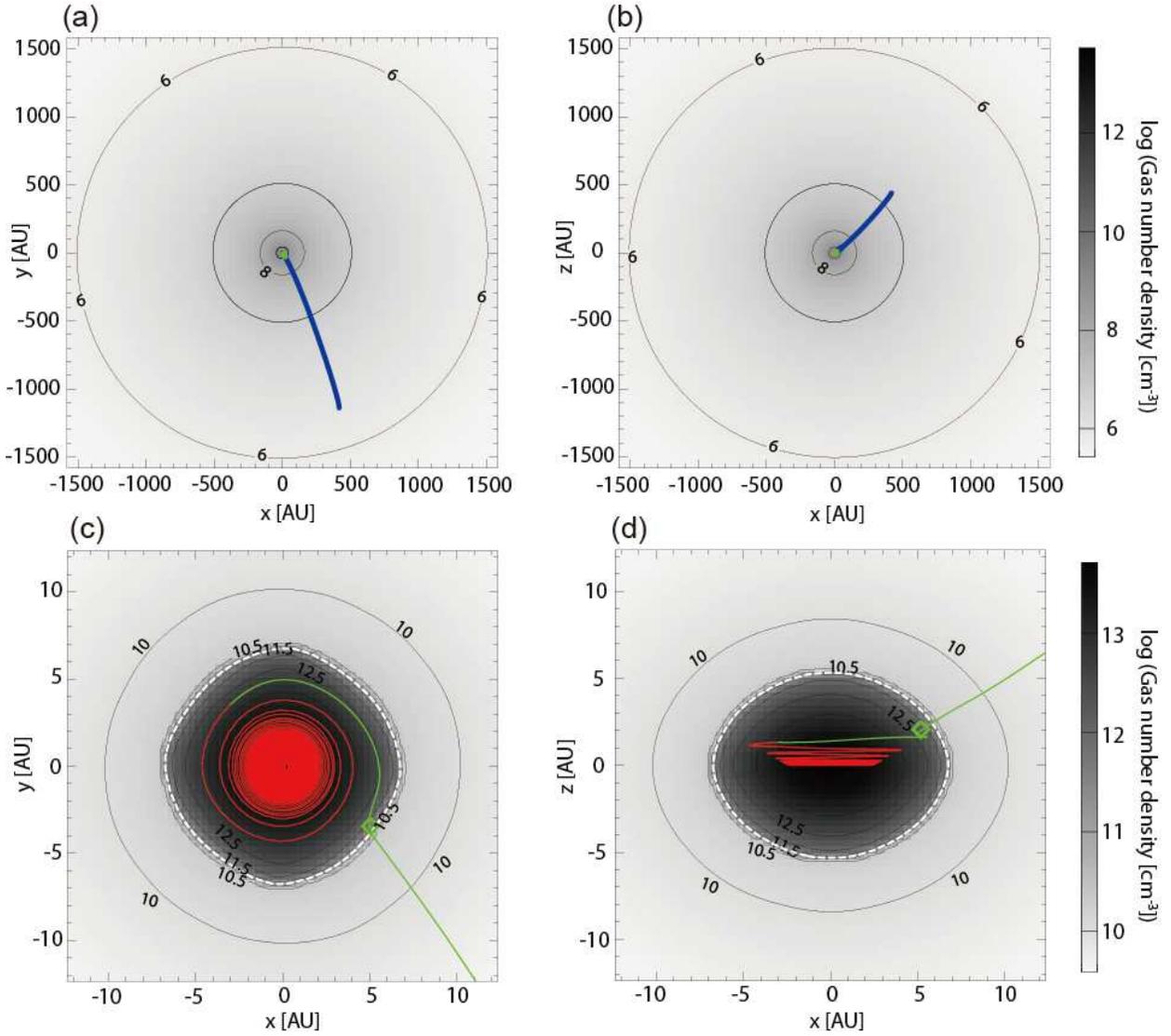}}
\caption{Trajectory of the fluid parcel, which is initially at 1300 AU from the center and 70$^\circ$ degrees from the rotational axis, plotted over the distribution of the gas number density when the parcel enters the first core (gray scales and contour lines). The values on the contour lines are number density in logarithm scale. The panels (a) and (b) show the whole trajectory, while the panels (c) and (d) are the closeup views. The trajectory is projected to the $z = 0$ plane in the panels (a) and (c), and to the $y = 0$ plane in the panels (b) and (d). The trajectory is color-coded according to the temperature in the parcel; the cold phase (10--20 K, blue), the warm-up phase (20--200 K, green) and the hot phase ($>$200 K, red). In the panels (c) and (d), the white dashed lines depict the first core surface, and the diamonds depict where the parcel passes the accretion shock at the first core surface. \label{fig5}}
\end{figure}

\begin{figure}
\epsscale{0.8}
\plotone{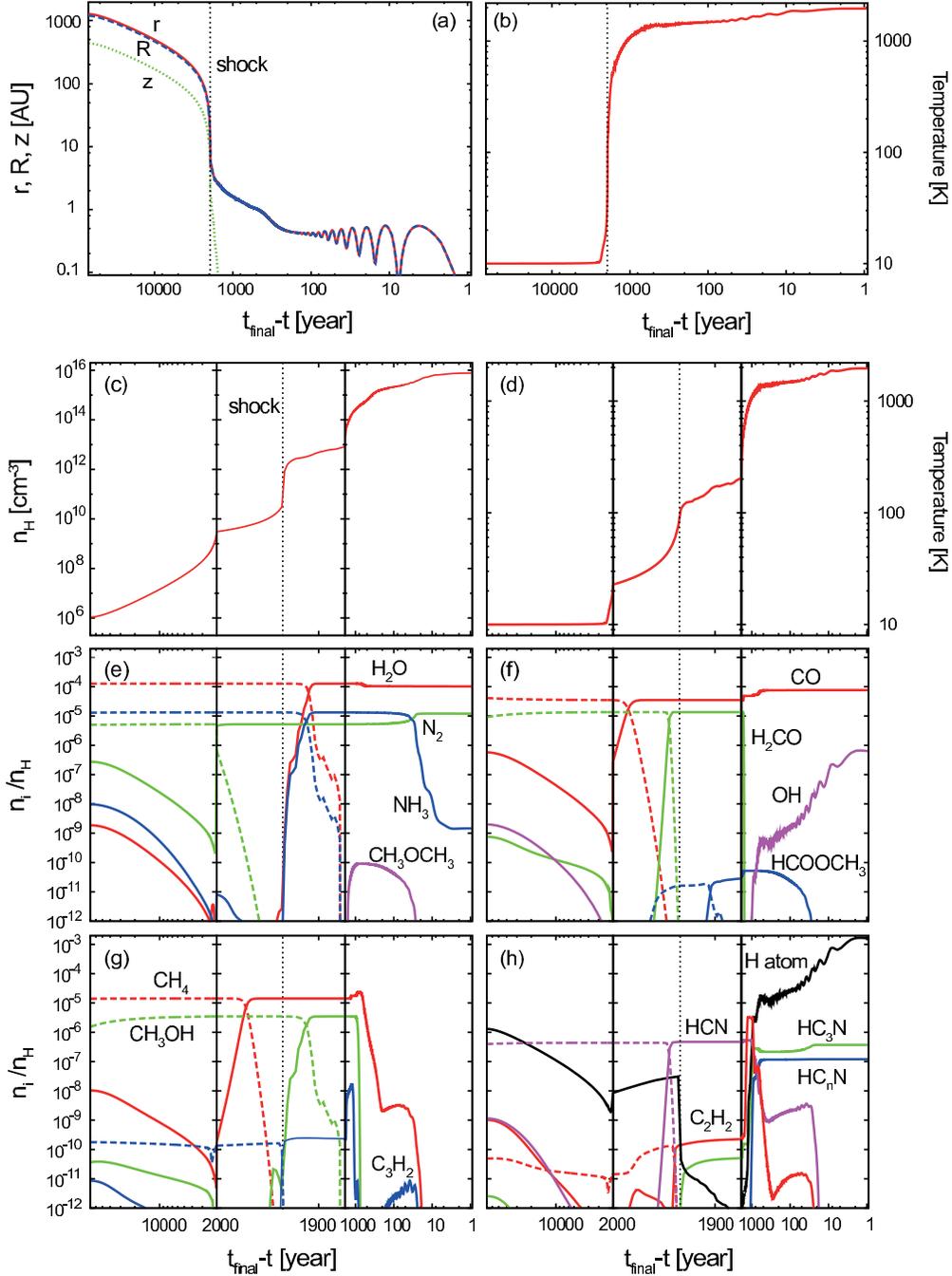}
\caption{Temporal variations of the position of the parcel in $r, R, z$ (a), temperature (b, d), number density of H nuclei (c), and molecular abundances (e--h) in the fluid parcel. The horizontal axis is the logarithm of $t_{\rm final}-t$. In order to zoom-up the warm-up phase, the panels (c--h) are divided into three phases; the cold phase ($T < 20$ K), the warm-up phase (20 K $< T < 200$ K) and the hot phase ($T>200$ K). Note that the scale of the horizontal axis is different among the phases. The vertical dotted lines represent the time when the parcel passes the accretion shock at the first core surface. In the panels (c--h), the solid lines represent gas-phase species, while the dashed lines represent ice-mantle species. HC$_n$N indicates total molecular abundances of HC$_5$N, HC$_7$N and HC$_9$N.\label{fig6}}
\end{figure}

\begin{figure}
\rotatebox{90}{\plotone{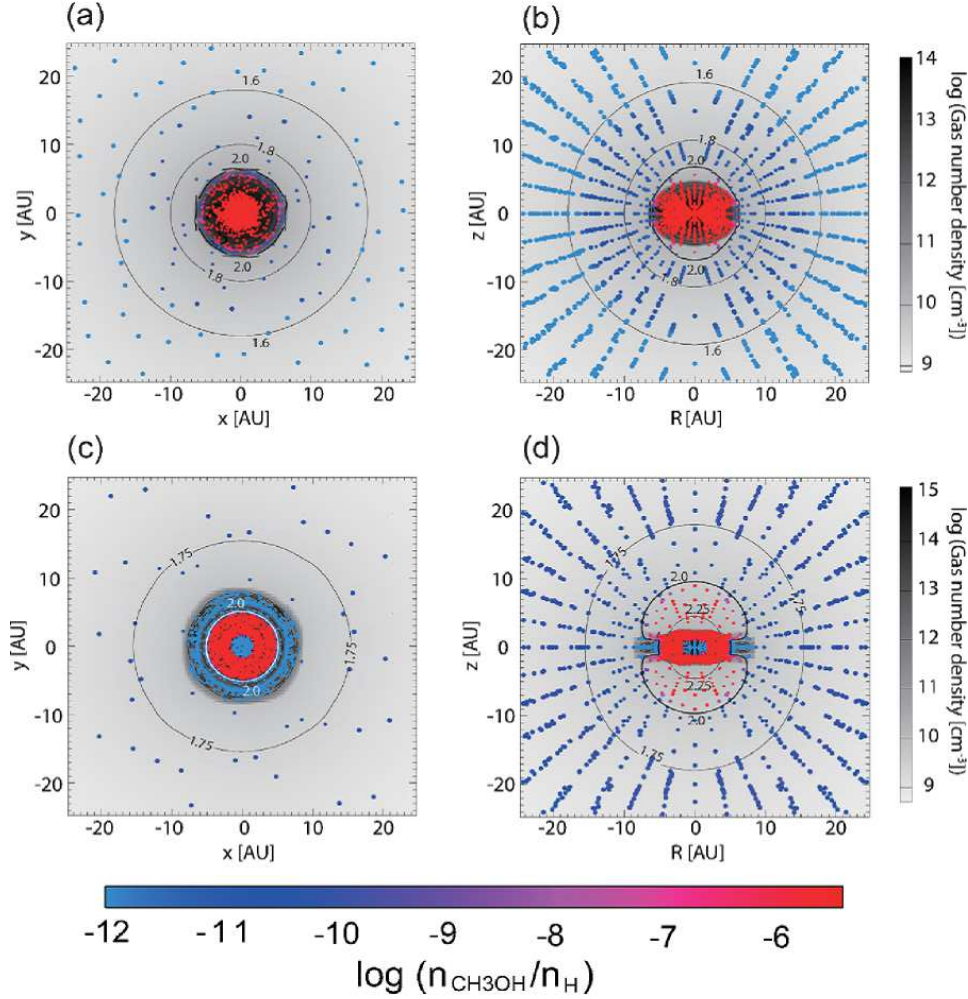}}
\caption{Spatial distributions of fluid parcels, which are color-coded according to abundance of gaseous CH$_3$OH, in the middle stage (a, b) and the late stage (c, d). The panels (a) and (c) represent the cross sections at $z = 0$, while (b) and (d) represent the $R$-$z$ plane. In the panels (a) and (c), the parcels with $|z_p| < 0.5$ AU are plotted. The gray scales and the contour lines represent the gas number density and temperature distributions, respectively in logarithmic scale.\label{fig7}}
\end{figure}

\begin{figure}
\epsscale{0.7}
\plotone{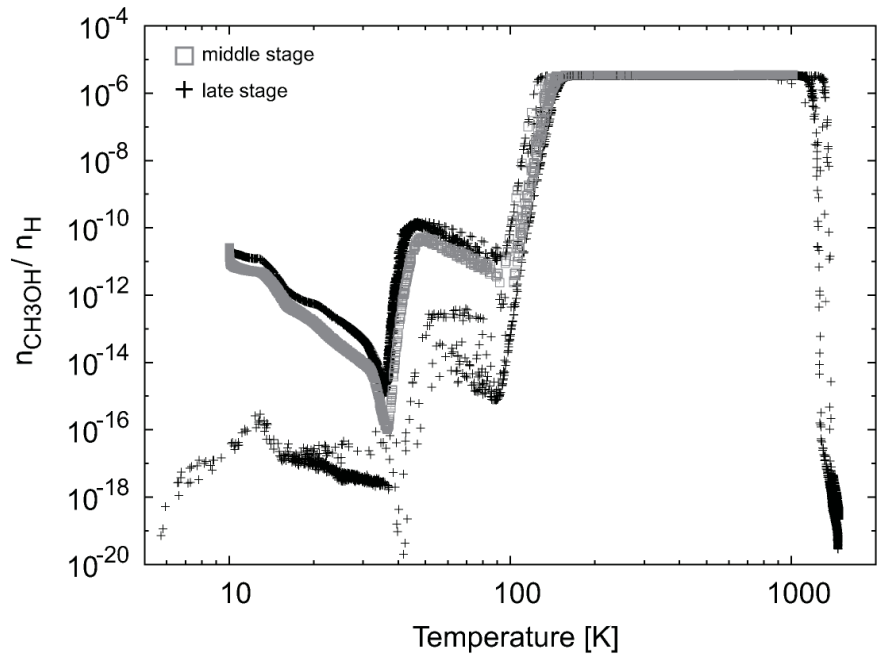}
\caption{Abundance of gaseous CH$_3$OH as a function of temperature in the middle stage (gray square) and the late stage (black cross).\label{fig8}}
\end{figure}

\begin{figure}
\epsscale{0.9}

\vspace{-1cm}

\plotone{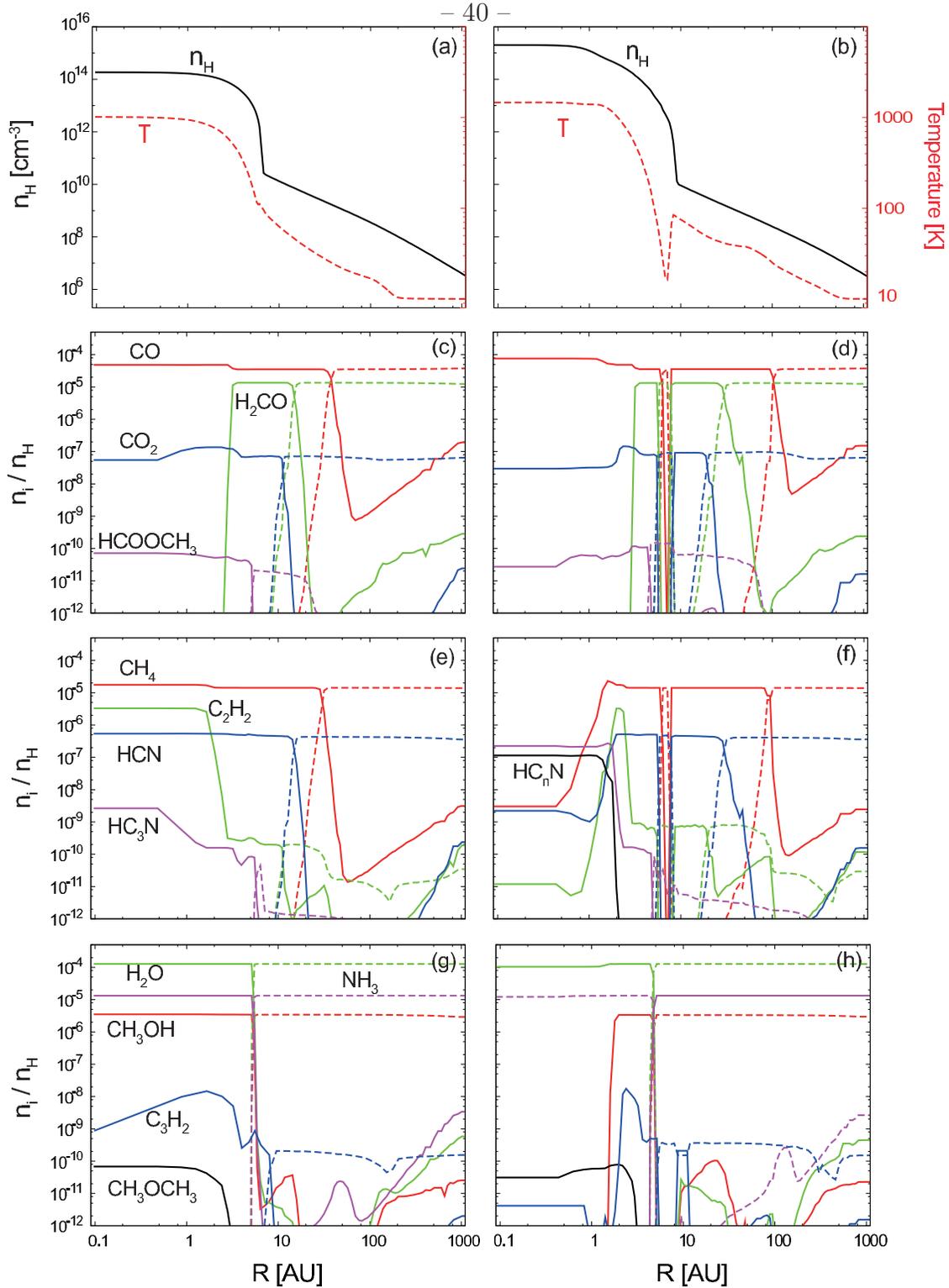}
\caption{Radial distributions of physical parameters (a, b) and molecular abundances (c--h) in the $z=0$ plane in the middle stage (a, c, e, g) and the late stage (b, d, f, h). In the panels (c--h), the solid lines represent gas-phase species, while the dashed lines represent ice-mantle species. HC$_n$N indicates total molecular abundances of HC$_5$N, HC$_7$N and HC$_9$N. \label{fig9}}
\end{figure}

\begin{figure}
\epsscale{0.9}

\vspace{-1cm}

\plotone{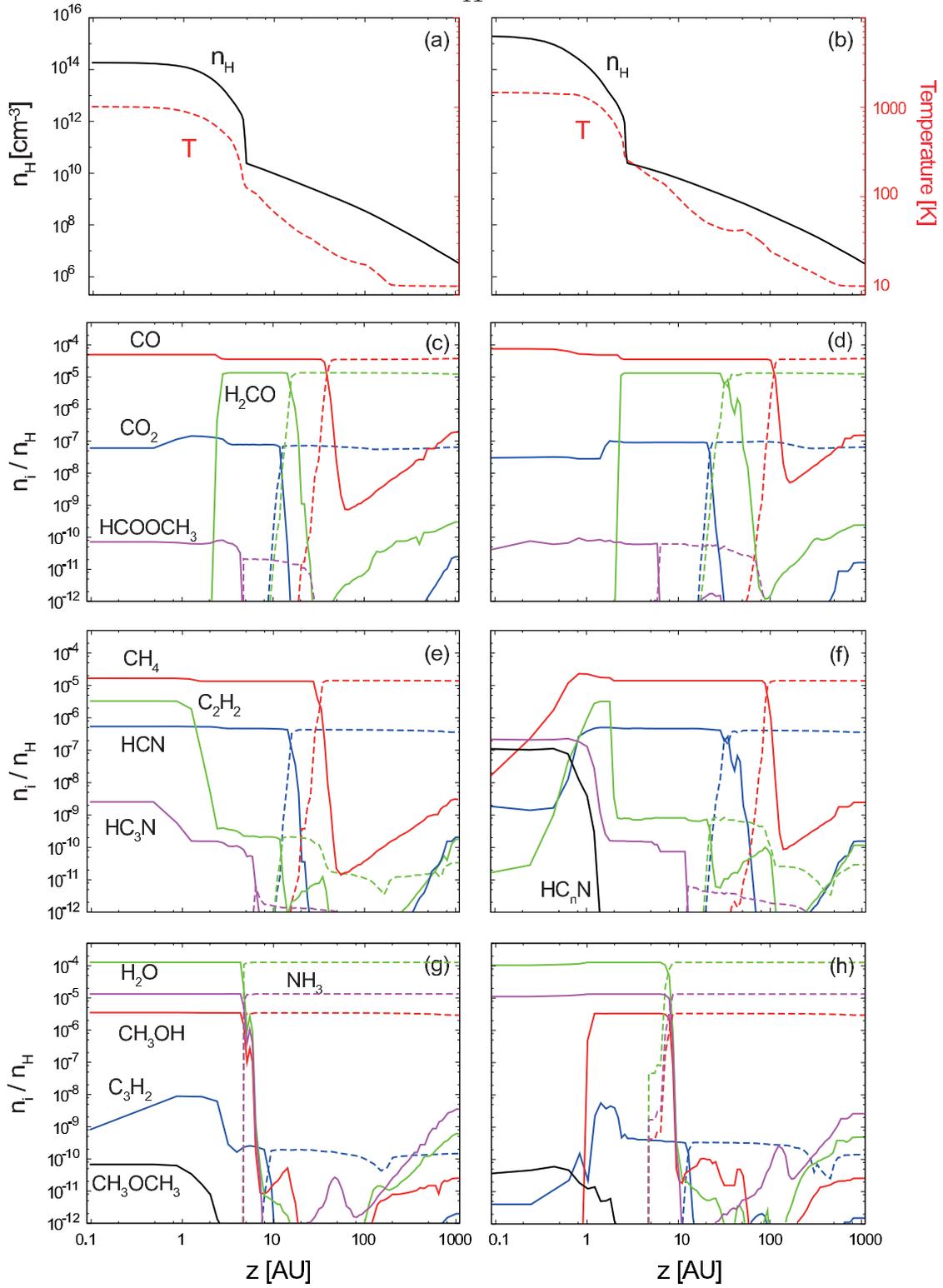}
\caption{Distributions of physical parameters (a, b) and molecular abundances (c--h) along the rotational axis. Other details are the same as in Figure \ref{fig9}.\label{fig10}}
\end{figure}

Figure \ref{fig9} shows the radial distributions of the physical parameters (a, b) and molecular abundances (c--h) in the $z = 0$ plane in the middle stage (a, c, e, g) and the late stage (b, d, f, h). The distributions along the rotational axis are shown in Figure \ref{fig10}, which is available in the online version of this paper. Since the temperature distributions are almost spherical (Figures \ref{fig2}(c) and \ref{fig2}(f)), and since the molecular abundances depend mainly on temperature, the molecular distributions are similar in Figure \ref{fig9} and Figure \ref{fig10}, except near the first core ($r \lesssim$ 10 AU). In the remainder of this subsection, we discuss the abundance profiles in detail.

\subsubsection{Simple Molecules}
Simple molecules are often observed both in prestellar and protostellar cores, and are used to probe the physical conditions. Among the neutral species observable in radio, CO has the lowest binding energy, and thus its sublimation radius is relatively large; $r \sim 40$ AU in the middle stage and $r \sim 100$ AU in the late stage. It could be observed as a CO emission spot inside the heavy CO depletion zone. Sublimation radii of H$_2$CO and HCN are also relatively large; $r \sim 15$ AU in the middle stage and $r \sim 30$ AU in the late stage. In the late stage, there is the low-temperature region ($T \sim 10$ K) inside the first core on the equatorial plane (Section 3.1). All gaseous species including CO are on grain surfaces there, because of the low temperature and high density. The freeze-out time scale is estimated to be only hours. Sublimation radii in the equatorial plane of the polar species, such as H$_2$O and NH$_3$ are small: $R \sim 5$ AU. Since the temperature distribution deviates from the spherical system in the small scale ($r \lesssim 10$ AU), the sublimation radii of the polar species are slightly larger along the rotational axis; $z \sim 8$ AU. In the middle stage the abundance of CH$_4$ is constant, but in the late stage the abundance drops at $r \sim 1$ AU. The peak abundances of species mentioned above do not change much with time.

The dominant positive ion is H$_3^+$ outside the sublimation radius of CO. At the inner radii, the dominant positive ion varies in accordance with proton affinity; HCO$^{+}$ dominates inside the CO sublimation radius, but it is replaced by H$_3$CO$^+$ and then NH$_4^+$, as H$_2$CO and NH$_3$ sublimate subsequently. The ionization degree drops toward the first core, due to increasing density. Dominant charge carrier is the charged grains at $n \gtrsim 10^{10}$ cm$^{-3}$.

\subsubsection{Carbon-Chain Molecules}
Unsaturated carbon-chain molecules are typically associated with cold dense cloud cores, but have also been detected toward the Class-0/I sources, L1527 \citep{sakai08} and IRAS 15398-3359 \citep{sakai09}. \citet{sakai08} suggested that carbon chains in the protostellar cores are formed from the sublimated CH$_4$ in the lukewarm region near the protostar. CH$_4$ reacts with carbon ion to form C$_2$H$_3^+$, which is a precursor of larger unsaturated carbon-chains. This formation scenario of carbon chains in the protostellar cores was named warm carbon-chain chemistry (WCCC) by \citet{sakai08}, and confirmed by AW08 and \citet{hassel08}.

We found WCCC does not efficiently occur in the first core stage; unsaturated carbon-chains do not increase at sublimation radii of CH$_4$, $r \sim 90$ AU and 30 AU in the middle and late stages. The abundance of carbon ion is lower due to the higher density of the envelope than in protostellar stages. Instead high temperature chemistry produces C$_2$H$_2$ and cyanopolyynes at $r <$ several AU. In the late stage, the abundance of C$_2$H$_2$ drops at $r \lesssim$ 1 AU, where longer cyanopolyynes, HC$_{\rm n}$N (n = 5, 7, 9), increase.

\subsubsection{Large Organic Molecules}
Large organic molecules sublimate at several AU. The total of gas and ice abundances of each species are almost constant throughout the core and envelope except for HCOOCH$_3$ and CH$_3$OCH$_3$. HCOOCH$_3$ is partly formed on grain surfaces at the lukewarm region, which expands from several tens AU (middle stage) to $\sim 100$ AU (late stage). CH$_3$OCH$_3$ is mainly formed in the gas phase at $r \lesssim 2$ AU. In the central region, the abundances of CH$_3$OH, HCOOCH$_3$ and CH$_3$OCH$_3$ drop in the late stage.






\section{DISCUSSION}
\subsection{Molecular Column Densities}
Detection and observation of first cores is one of the most important challenges in ALMA era. First cores are characterized by a AU-size hydrostatic core and low-velocity outflow with a wide opening angle. While size of the core can be measured by high angular resolution dust continuum observation, molecular line observations are also necessary to confirm the kinetic structure. Molecular abundances obtained in our model are useful to predict where the various molecular lines would arise. We especially focus on large organic species, which are abundant at the vicinity and the surface of the first core. Since detailed radiation transfer modeling \citep[e.g.,][]{tomisaka11} is out of the scope of this work, here we compare the column densities of assorted species in our model with those estimated from the observations towards a prototypical hot corino, IRAS 16293-2422, to evaluate if our model column densities of large organic species are high enough to produce lines.

\begin{figure}
\epsscale{0.7}
\plotone{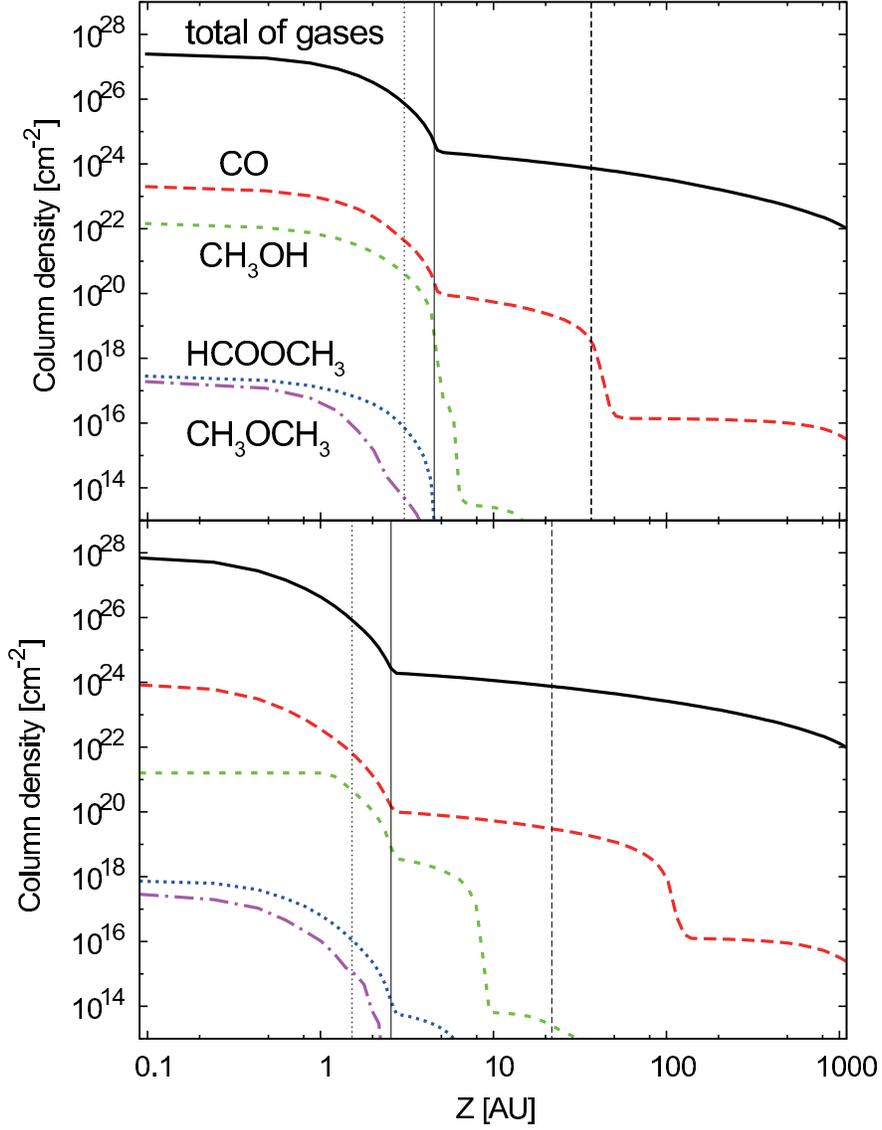}
\caption{Column densities along the rotational axis integrated from the outer edge of the envelope to $z$, $N_{z, i}(z) = \int_{z}^{\infty}n_idz'$, in the middle stage (upper panel) and the late stage (lower panel). The vertical solid lines represent the first core surface. The vertical dashed and dotted lines represent the heights from the midplane, where the optical depths of dust continuum reach unity with the wave length of 100 $\mu$m and 1 mm, respectively. \label{fig11}}
\end{figure}

\begin{figure}
\epsscale{0.7}
\plotone{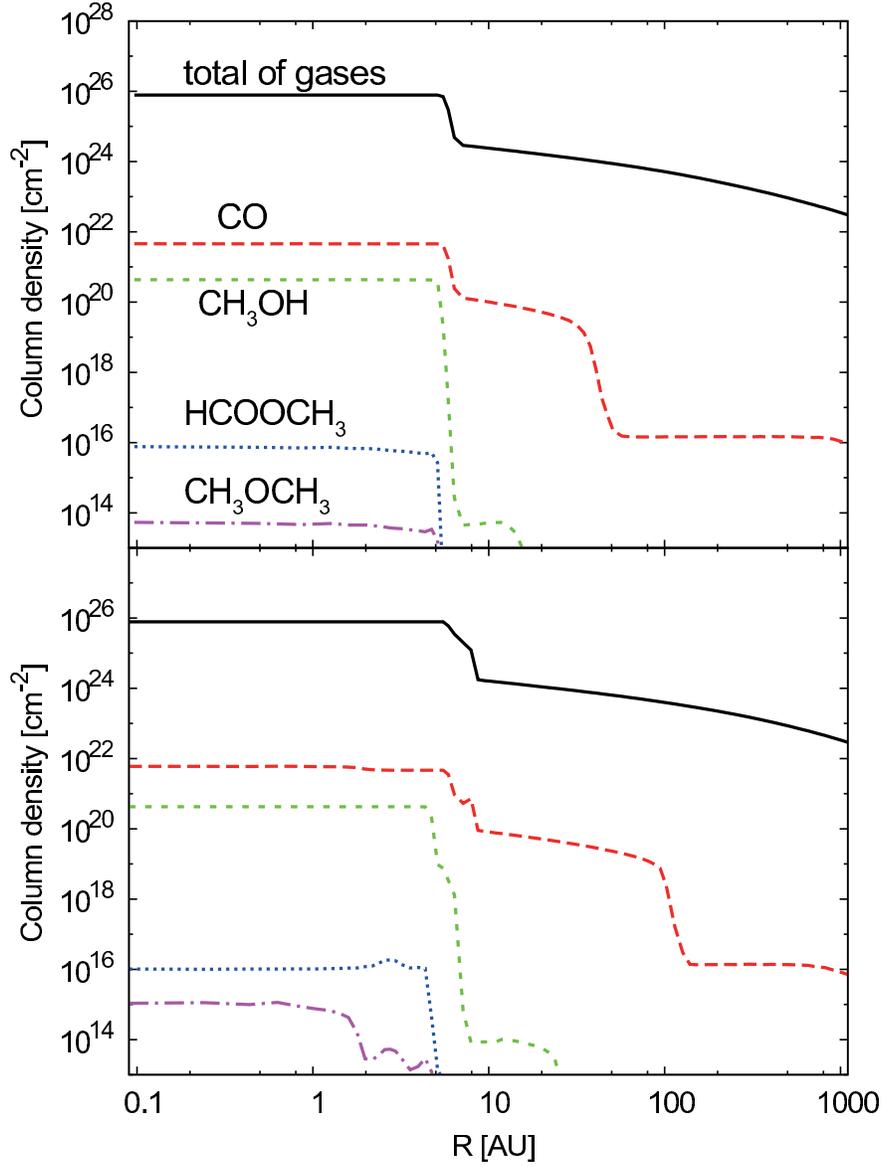}
\caption{Gaseous molecular column densities in front of the $\tau_{\rm {1 mm}} = 1$ plane in the face-on object. \label{fig12}}
\end{figure}

Figure \ref{fig11} (solid black line) shows the cumulative column densities of gases along the rotational axis from the outer edge of the core to a height $z$ above the midplane, $N_z(z) = \int_{z}^{\infty}ndz'$. The vertical solid lines represent the first core surface. The vertical dashed and dotted lines represent the heights where the optical depths of dust continuum reach unity at the wavelength of 100 $\mu$m and 1 mm, respectively. The optical depth of dust continuum, $\tau$, is estimated as
\begin{equation}
\tau = 1\left(\frac{N_z}{8\times10^{25}\hspace{2pt}{\rm cm^{-2}}}\right)\left(\frac{\lambda}{1\hspace{2pt}{\rm mm}}\right)^{-2},
\end{equation}
referring to Hildebrand (1983). At 100 $\mu$m the optical depth reaches unity in the envelope. At 1 mm, on the other hand, we can look into the envelope and the surface of the first core, although the optical depth exceeds unity at deeper layers of the first core \citep[see also][]{saigo11}. The gases in front of the $\tau_{\rm 1 mm} \sim 1$ plane can be probed by absorption lines, since the gas in front is cooler than the $\tau_{\rm 1 mm} \sim 1$ plane. In the line of sight offset from the rotational axis where dust continuum is optically thin ($R \gtrsim 6$ AU), the line could be emission (Figure \ref{fig12}). We calculate the molecular column density in front of the $\tau_{\rm 1 mm} = 1$ plane in the face-on system,
\begin{equation}
N_{z_{\tau=1}, i}(x, y) = \int_{z_{\tau=1}(x, y)}^{\infty}n_i(x, y, z')dz',
\end{equation}
where $z_{\tau=1}$ represents the height $z$ where the optical depth reaches unity at 1 mm, and $n_i$ represents the number density of species $i$. Figure \ref{fig12} shows $N_{z_{\tau=1}, i}$ of assorted molecules as a function of $R$ in the middle stage (upper panel) and the late stage (lower panel).

At $R \lesssim 6$ AU the first core is optically thick for 1 mm dust continuum, and $N_{z_{\tau=1}, i}$ is almost constant. $N_{z_{\tau=1}, {\rm CO}}$ sharply increases inwards at its sublimation radius (several tens AU) and at the first core surface (several AU). $N_{z_{\tau=1}, i}$ of large organic molecules sharply increases at their sublimation radii which is close to the first core surface. The peak value of $N_{z_{\tau=1}, i}$ for each species are not much different between the middle and late stages. CH$_3$OCH$_3$ is an exception; its peak value of $N_{z_{\tau=1}, {\rm CH_3OCH_3}}$ increases with time.
 
The emission of these complex organic molecules have been detected towards several low mass protostars \citep[e.g.,][]{cazaux03,bottinelli04,maret05}. The beam-averaged column densities derived from interferometric observations toward IRAS 16293-2422 are $N_{\rm CH_3OH}=1.1\times10^{18}$ cm$^{-2}$, $N_{\rm HCOOCH_3}=6.8\times10^{15}$ cm$^{-2}$ \citep{kuan04}, and $N_{\rm CH_3OCH_3}=6.4\times10^{15}$ cm$^{-2}$ \citep{chandler05}. \citet{kuan04} also derived $N_{\rm ^{13}CH_3OH}=8.1\times10^{16}$ cm$^{-2}$. It indicates that the $^{12}$CH$_3$OH emission is optically thick, since the carbon isotope ratio of CH$_3$OH, $N_{\rm ^{12}CH_3OH}$/$N_{\rm ^{13}CH_3OH} \sim 14$, is much smaller than the elemental abundance ratio, $[^{12}{\rm C}/^{13}{\rm C}]\sim60$, in ISM \citep[e.g.,][]{langer93,lucas98}. Our model column densities of CH$_3$OH and HCOOCH$_3$ in the middle and late stages at $R \lesssim 6$ AU are comparable or larger than those values. Hence it would be reasonable to expect that the column densities of CH$_3$OH (and HCOOCH$_3$) at $R \lesssim 6$ AU in our model are high enough to be $\tau_l \sim 1$ at least in some transitions. Since the gas in the first core surface and the envelope is cooler than the first core, we expect absorption lines, if we observe the object from the pole-on angle. Assuming the lines and continuum are thermalized, the expected difference of brightness is
\begin{equation}
\Delta T_b = |T_{bg} - T_{fg}|[1-\exp(-\tau_l)],
\end{equation}
where $T_{bg}$ and $T_{fg}$ are the temperatures at $\tau_{1{\rm mm}}=1$ and $\tau_l=1$, respectively. We assume that $\tau_l$ for CH$_3$OH reaches unity at the height where the model column density of CH$_3$OH reaches $1.1\times10^{18}$ cm$^{-2}$. In the middle stage for CH$_3$OH, $\Delta T_b$ is $\gtrsim$40 K at $R\lesssim6$ AU, which is detectable in 1.6 hours toward the distance of 120 pc with S/N $\gtrsim$ 10 and velocity resolution of 0.1 km/s in the extended configuration ($\sim$0.1") of ALMA at Band 7 (345 GHz). More integration time would be required to detect HCOOCH$_3$, because the $\tau_l = 1$ plane for HCOOCH$_3$ is closer to the $\tau_{\rm 1 mm} = 1$ plane than that for CH$_3$OH is (i.e., $\Delta T_b$ for HCOOCH$_3$ is smaller) (see Figure \ref{fig11}). We found that the column densities of HCOOH and CH$_3$CN in our model are also comparable to the observed values in the hot corinos. Although, detailed radiation transfer calculation is needed to confirm these estimates.

On the other hand, the column density of CH$_3$OCH$_3$ in our model is much less than the observed toward IRAS 16293-2422 both in the middle and late stages, and thus would not be high enough to produce observable lines.




Our results show that first cores and the surrounding warm envelopes could be observed as very compact hot corinos without stellar signatures, and the large organic molecules can be good tracers of first cores. In the rest of this paper, we call the warm gases with the large organic molecules a prestellar hot corino.

\subsection{Uncertainties in the Reaction Network Model}
There are several caveats to our model abundances and column densities of large organic molecules presented above. As mentioned in Section 3.3.3, H$_2$CO is collisionally dissociated at lower temperatures ($\sim$500 K) than other large organic molecules, since the value of $\gamma$ ($1.75\times10^4$ K) is small for the collisional dissociation, H$_2$CO + M $\rightarrow$ H$_2$ + CO + M, compared with those of other large organic molecules. It should be noted that the value of $\gamma$ in the collisional dissociations listed in Table A2 are basically determined by the difference between the formation enthalpy of reactants and products, but the products and branching ratios for various possible sets of products are not well constrained by laboratory experiments (Section 2.2.1). If there exists the  channels with lower $\gamma$ for the destruction of other large organic molecules, the abundances of these molecules at $T \gtrsim 500$ K could decrease by several orders of magnitude relative to our results. The column densities of the large organic molecules in the middle stage (Figure \ref{fig12}, upper panel), however, would not change, since the temperature is lower than 500 K at $\tau_{\rm 1mm} \lesssim 1$. The column densities in the late stage (Figure \ref{fig12}, lower panel) would decrease by about one order of magnitude.

The abundances of HCOOCH$_3$ and CH$_3$OCH$_3$ in our model depend on the branching ratios of grain surface recombination, which contributes to their formation in the hot phase. We recalculate the molecular abundances in the representative fluid parcel discussed in Section 3.3, assuming all the product channels of grain surface recombination (both with a negatively charged grain and with a neutral grain) are equally weighted. Then, the peak abundances of HCOOCH$_3$ and CH$_3$OCH$_3$ are enhanced by a factor of 8 and 12, respectively. The column densities would be enhanced almost by the same factor, since grain surface recombination is effective around/in the dense first core. In contrast, the abundance of CH$_3$OH does not strongly depend on the assumed branching ratio of the recombination, since CH$_3$OH is mostly formed by hydrogenation on grain surfaces in the cold phase.

While the cosmic ionization rate is set to be $1.3\times10^{-17}$ s$^{-1}$ in our model, a larger value of $5\times10^{-17}$ s$^{-1}$ is suggested by recent work \citep{dalgarno06}. We recalculate the molecular abundances in the representative fluid parcel with the cosmic-ray ionization rate of $5\times10^{-17}$ s$^{-1}$. The peak abundances of large organic molecules are enhanced by a factor of 2--5 compared to our original model with the cosmic-ray ionization rate of $1.3\times10^{-17}$ s$^{-1}$.

In summary, our model might underestimate the abundances of large organic molecules at $T \lesssim 500$ K by one order of magnitude, while our model might overestimate them at $T \gtrsim 500$ K by several orders of magnitude. The model column densities might be underestimated in the middle stage by one order of magnitude, while those in the late stage could be higher or lower by one order of magnitude.

\subsection{Comparison with Protostellar Hot Corinos}
Here, we compare the physical structures and chemical compositions of the prestellar hot corino (prestellar HC) with those of a protostellar hot corino (protostellar HC). We refer to the spherical RHD model of a star-forming core by \citet{masunaga00}, and AW08 for the protostellar hot corino model. Such comparison is useful to unveil differences between prestellar and protostellar HCs. It should be noted, however, that the physical structures of the cores are different; e.g., while the models of \citet{masunaga00} and AW08 are non-rotating cores of 3.9 $M_{\bigodot}$, our model is the rotating core of 1 $M_{\bigodot}$.

The top two columns of Table 4 show the mean radius and the gas number density of $T = 100$ K region. It is clear that the prestellar HC is more compact and denser than the protostellar HC by an order of magnitude; the protostar is more luminous and their envelopes are more sparse than in the first core stage.

We compare the abundances of the large organic molecules in the central region ($r \sim 5$ AU) of our model with those of the protostellar HC model. While AW08 traced 13 parcels in the RHD simulation of \citet{masunaga00} to calculate the molecular abundances, the reaction network used in AW08 is different from our network. In order to make comparisons easy, we revisit the calculation of AW08 with some modifications; we calculate the temporal variations of the molecular abundances using our network model in the parcel which reaches $r = 15$ AU at the protostellar age of $9.3\times10^4$ yr. We adopt this parcel, because the abundances are mostly constant at $r \lesssim 100$ AU according to AW08. While AW08 used the initial temperature of 6 K, we set it at 10 K when the temperature is lower than 10 K in the original model. Table 4 shows the abundances of the large organic molecules at 15 AU in the protostellar HC model and 5 AU in the prestellar HC model. Comparing the second row of Table 4 with Table 3 of AW08, our modifications in the network and temperature lead to order-of-magnitude changes in the abundances of listed species except for HCOOCH$_3$. Among the listed species, the abundances of H$_2$CO and CH$_3$OH are similar in the prestellar HC model and the protostellar HC model. These species are formed during the cold phase by the subsequent hydrogenation reactions of CO on grain surfaces. The other species are less abundant in the prestellar HC model. In the protostellar HC model, they are efficiently formed in the lukewarm region. On the other hand, in the prestellar HC model the lukewarm region is small, and the time to form these species is limited. It indicates that the abundances of these molecules are sensitive to the luminosity of the central object.

\subsection{Accretion Shock Heating Layers}
The supersonic accretion of the envelope material to the first core causes the shock wave. This accretion shock is isothermal, because the radiation from post-shock gas heats up pre-shock gas, and the temperature in the pre-shock and post-shock gases become equal \citep[e.g.,][]{omukai07,saigo11,commercon11}. Between the pre- and post-shock gases, the shock heating layer is formed, in which the gas temperature reaches $\sim$2000 K at the maximum. In contrast, the grain temperature increases to only $\sim$200 K. The thickness of this layer is $\sim$10$^{-3}$ AU, and the cooling time scale of the gas is $\sim$1 hour after crossing the shock front \citep[see Fig. 3 of][]{saigo11}. In our hydrodynamic simulation the shock heating layer is not spatially resolved, and thus our chemical calculations neglect the temperature rise in the shock heating layer. In order to evaluate the effect of shock heating on chemistry, we recalculate the molecular abundances in the parcel discussed in Section 3.3 by adding 1 hour integration at the constant temperature when the parcel reaches the first core surface. Since the temperature of the shock heating layer should vary depending on the mass of the first core and the radius where the shock occurs, we calculate two cases: the gas temperature of 1000 K and 2000 K. The density is assumed to be the same as the post-shock gas along the trajectory, $n_{\rm H} = 8.5 \times 10^{11}$ cm$^{-3}$. The grain temperature is assumed to be 200 K, at which temperature ice-mantle species are almost evaporated. In the pre-shock and post-shock regions, on the other hand, the species with the high binding energies, such as CH$_3$OH, is mostly freeze-out on grain surfaces, since the grain temperature is about 80 K. To focus on high temperature chemical processes, we discuss changes of the total (gas and ice combined) abundances in this subsection.

\begin{figure}
\epsscale{0.7}
\plotone{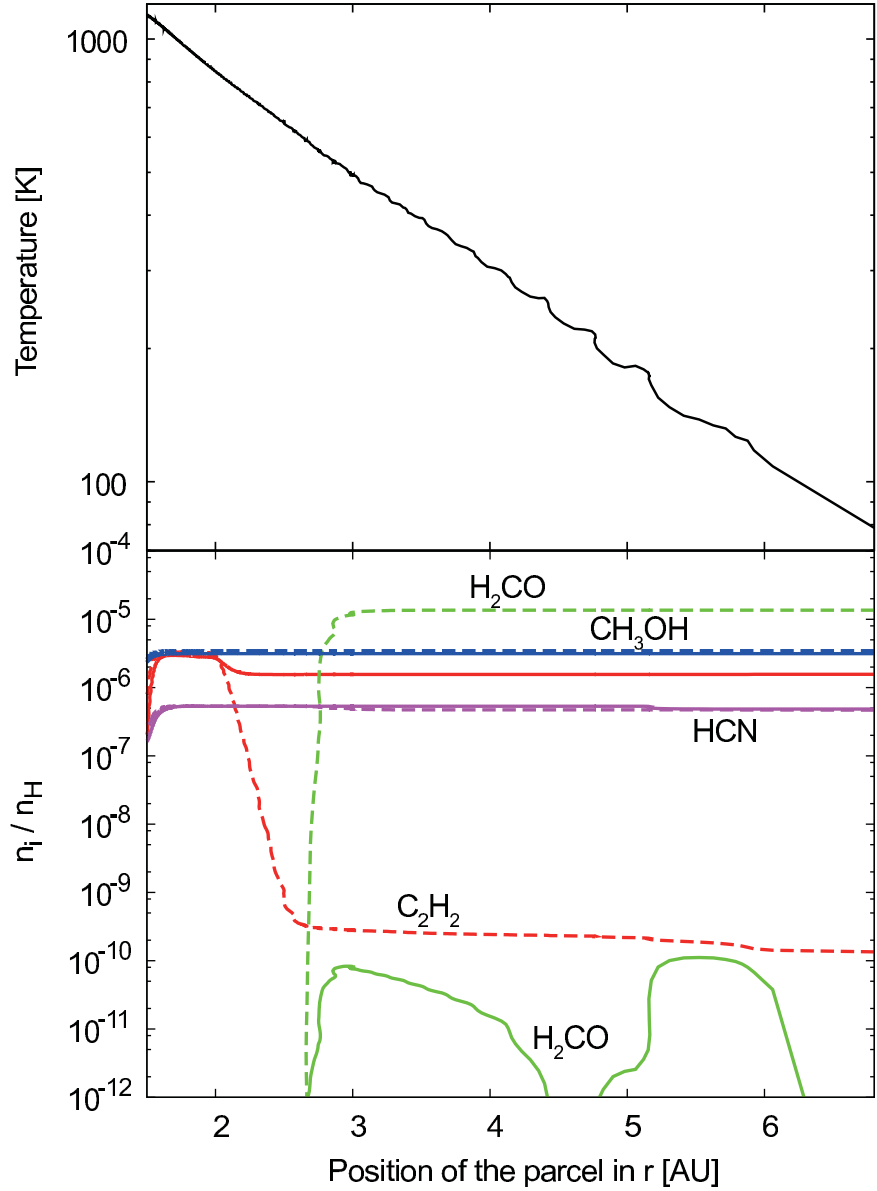}
\caption{Temperature (upper panel) and total (gas and ice) molecular abundances of selected species (lower panel) along the trajectory discussed in Section 3.3 after the parcel enter the first core at $r \sim 7$ AU. The solid and dashed lines in the lower panel represent the abundances with and without the shock heating, respectively. \label{fig13}}
\end{figure}

In the case of the gas temperature of 1000 K, the shock heating does not affect the chemistry. Even in the case of the gas temperature of 2000 K, the heating has a limited effect on the chemistry; among abundant species ($n_{i}/n_{\rm H} > 10^{-12}$), only 23 species increase or decrease by a factor of $>$2 in the shock heating layer. Figure \ref{fig13} shows the spatial variation of the temperature (upper panel) and molecular abundances (lower panel) along the trajectory after the parcel experience the shock heating of 2000 K at $r \sim 7$ AU and $t_{\rm final}-t=1930$ yr (see Figure \ref{fig6}). The horizontal axis in Figure \ref{fig13} is the position of the parcel in $r$. In the lower panel, the solid lines show the molecular abundances with shock heating, while the dashed lines show our original model without the shock heating. The abundances of HCN and CH$_3$OH are almost the same between the two models. H$_2$CO is the most sensitive to the shock heating; it decreases approximately five orders of magnitude by collisional dissociation to form CO and H$_2$. Since the abundances of CO and H$_2$ are higher than that of H$_2$CO, the effect of H$_2$CO dissociation to other molecules is insignificant. The difference between the two models diminishes when the temperature reaches $\sim$500 K, at which collisional dissociation of H$_2$CO becomes efficient in the original model. Carbon species, such as C$_2$H$_2$, are also sensitive to the shock heating; their abundances are temporally enhanced by approximately four orders of magnitude by the reaction of C$_3$H$_4$ and H$_2$ with the potential barrier of $2.3\times10^4$ K. The difference in the C$_2$H$_2$ abundance diminishes when the temperature reaches $\sim$800 K. Therefore our results in Section 3 might overestimate the abundance of H$_2$CO, while underestimate the abundance of C$_2$H$_2$ in the first core except for the hot central region ($r \lesssim 3$ AU). 

While our simple calculation shows that the shock heating affects abundances of some molecules, the effect is limited and the shock heating does not lead to a significant change in our results presented in Section 3.

\subsection{Toward the Chemistry in Circumstellar Disks}
The first core is a transient object formed prior to the protostar. Here, we briefly discuss what happens after the formation of the protostar and the circumstellar disk in terms of chemistry.

There are several studies which investigate the formation and evolution of the circumstellar disk using three-dimensional hydrodynamic simulations. We refer to \citet{machida10,machida11}, which investigated a longest-term evolution of the circumstellar disk, starting from the gravitational collapse of the molecular cloud core. While the central part ($r \lesssim 1$ AU) of the first core collapses to form a protostar, the outer part directly evolves to the cicumstellar disk with a Keplerian rotation. In that case, the chemical composition of the first core corresponds to the initial composition of the circumstellar disk; the total of gas and ice abundances at $r \gtrsim 3$ AU ($T \lesssim 500$ K) in the first core is mostly the same as those in the cold molecular cloud core, while the inner part ($T \gtrsim 500$ K) is dominated by the simple molecules, such as CO, H$_2$O and N$_2$.

When the protostar is born, the mass of the circumstellar disk ($\sim$0.01--0.1 $M_{\bigodot}$) overwhelms that of the protostar ($\sim$10$^{-3}$ $M_{\bigodot}$). It causes the gravitational instability, and angular momentum is efficiently transported via spiral arms. The mass of the protostar and circumstellar disk increase with time via the accretion from the circumstellar disk and infalling envelope, respectively. Therefore most of the material constituting the first core would accrete to the protostar, and be replaced by the newly accreted envelope material during the disk evolution, although a small fraction of the first core material may have been transported to outer region of the disk carrying angular momentum. Since the envelope material in the protostellar stage is more abundant in large organic molecules than in the first core stage (Section 4.3), the circumstellar disk would be enriched with the large organic molecule. The large organic molecules also might be formed in the circumstellar disk via gas-phase and/or grain-surface reactions if the fluid parcels stay in the disk long enough ($10^4$--$10^5$ yrs) (e.g., Charnley et al. 1992; Garrod \& Herbst 2006; Aikawa et al. in prep). On the other hand, increasing UV radiation and decreasing attenuation in the envelope cause the photodissociation of molecules, such as H$_2$O and CO, at the surfaces of the disk and the wall of the outflow cavity \citep[e.g.,][]{visser11}.

\section{CONCLUSION}
We have investigated molecular evolution in the first hydrostatic core stage. We performed the three-dimensional radiation hydrodynamic simulation from the molecular cloud core to the first core. We traced the fluid parcels, which follow the local flow of the gas, and solved molecular evolution along the trajectories, coupling gas-phase and grain-surface chemistry. We derived the spatial distributions of molecular abundances and column densities in the molecular cloud core harboring the first core. Our conclusions are as follows.

1. We found that the total molecular abundances of gas and ice of many species in the cold era (10 K) remain unaltered until the temperature reaches  $\sim$500 K. The gas abundances in the warm envelope and outer layers of the first core ($T \lesssim 500$ K) are mainly determined via the sublimation of ice-mantle species. Above 500 K, the abundant molecules, such as H$_2$CO, start to be destroyed mainly via collisional dissociation or reactions with H atom, and converted to the simple molecules, such as CO, H$_2$O and N$_2$. On the other hand, some molecules are effectively formed; carbon-chains, such as C$_2$H$_2$ and cyanopolyynes, are formed at the temperature of $>$700 K.

2. Gaseous large organic molecules are associated with the first core ($r < 10$ AU). Although the abundances of large organic molecules in the first core stage are comparable or smaller than in the protostellar hot corino, their column densities in our model are comparable to the observed values toward the prototypical hot corino, IRAS 16293-2422. We propose that large organic molecules can be good tracers of the first cores. 

3. The shock wave occurs when the envelope material accretes onto the first core. We found that the shock heating has little effect on chemistry, if the maximum temperature in the shock heating layer is below 1000 K. Even if the gas temperature reaches 2000 K, the heating has a limited effect on chemistry. While it significantly decreases the H$_2$CO abundance and increases the C$_2$H$_2$ abundance in the outer parts of the first core ($r \lesssim 3$ AU), it has little effect on other species.

4. Recent hydrodynamic simulations have shown that outer region of the first core could directly evolve to the circumstellar disk. In that case, the chemical composition of the first core corresponds to the initial composition of the circumstellar disk; the total abundances of gas and ice at $r \gtrsim 3$ AU ($T \lesssim 500$ K) are mostly determined in the cold molecular cloud cores, while inner parts ($T \gtrsim 500$ K) are dominated by simple molecules, such as CO, H$_2$O and N$_2$.




\acknowledgments
Acknowledgements. We thank Ms. Audrey Coutens and Professor Toyoharu Umebayashi for useful discussions. We also thank the anonymous referee for the helpful comments that improved the manuscript. This work is partly supported by Grant-in-Aids for Scientific Research, 21244021 (K. Tomisaka and Y. Aikawa), 23103004, 23540266 (Y. Aikawa), and 23540270 (T. Matsumoto), and Global COE program "Foundation of International Center for Planetary Science" (G11) of the Ministry of Education, Culture, Sports, Science and Technology of Japan (MEXT). K. Furuya and K. Tomida are supported by the Research Fellowship from the Japan Society for the Promotion of Science (JSPS) for Young Scientists. F. Hersant and V. Wakelam are grateful to the French CNRS/INSU PCMI for its financial support. Numerical computations were partly carried out on NEC SX-9 at Japan Aerospace Exploration Agency (JAXA) and at Center for Computational Astrophysics (CfCA) of National Astronomical Observatory of Japan. Some kinetic data we used have been downloaded from the online database KIDA (http://kida.obs.u-bordeaux1.fr).

\clearpage

\clearpage

\clearpage

\begin{table}
\begin{center}
\caption{Generalized reaction scheme of gas-grain interactions. X, Y, and Z are neutral gaseous species whose charges are denoted by their indices, and gr are grain particles. X ice represents the species on grain surfaces.}
\begin{tabular}{lccccccccc}
\tableline\tableline
Grain-grain &gr$^-$ & + & gr$^+$ & $\rightarrow$ & gr     & + & gr     &   &\\
\tableline
Electron-grain&e$^-$  & + & gr     & $\rightarrow$ & gr$^-$ &   &        &   &\\
&e$^-$  & + & gr$^+$ & $\rightarrow$ & gr     &   &        &   &\\
\tableline
Cation-grain&X$^+$  & + & gr$^-$ & $\rightarrow$ & Y      & + & Z      & + & gr \\
&X$^+$  & + & gr     & $\rightarrow$ & Y      & + & Z      & + & gr$^+$ \\
\tableline
Anion-grain&X$^-$  & + & gr$^+$ & $\rightarrow$ & X      & + & gr     &   & \\
\tableline
Adsorption&X      & + & gr$^-$ & $\rightarrow$ & X ice  & + & gr$^-$ &   &\\
&X      & + & gr     & $\rightarrow$ & X ice  & + & gr     &   &\\
&X      & + & gr$^+$ & $\rightarrow$ & X ice  & + & gr$^+$ &   &\\
\tableline
Desorption&X ice &   &        & $\rightarrow$ & X      &   &        &   &\\
\tableline
\end{tabular}
\end{center}
\end{table}

\begin{table}
\begin{center}
\caption{Initial abundances relative to hydrogen nuclei.}
\begin{tabular}{cc}
\tableline\tableline
Species & Abundance\footnotemark[1] \\
\tableline
H$_2$.............................................. &5.00(-1)\\
H............................................... &5.00(-5)\\
He.............................................. &9.75(-2)\\
C$^+$.............................................. &7.86(-5)\\
N............................................... &2.47(-5)\\
O............................................... &1.80(-4)\\
Si$^+$............................................. &2.74(-9)\\
S$^+$.............................................. &9.14(-8)\\
Fe$^+$............................................. &2.74(-9)\\
Na$^+$............................................. &2.25(-9)\\
Mg$^+$............................................. &1.09(-8)\\
Cl$^+$............................................. &1.00(-9)\\
P$^+$.............................................. &2.16(-10)\\
\tableline
\end{tabular}
\tablenotetext{1}{$a(-b)$ means $a\times10^{-b}$.}
\end{center}
\end{table}

\renewcommand\thefootnote{\alph{footnote}} 

\begin{table}
\begin{center}
\caption{Temperature, number density of hydrogen nuclei, position in $r$, and duration of the cold, warm-up, and hot phases in the fluid parcel. $t_{ff}$ is the free-fall time of the initial central density.}
\footnotesize
\begin{tabular}{ccccc}
\hline\hline
 & Temperature & $n_{\rm H}$ & Position in $r$ & Duration \\
 & (K) & (cm$^{-3}$) & (AU) & (yr) \\
\hline
Cold phase & 10--20 & $1\times10^6$--$2\times10^{9}$ & 1500--40 & $7\times10^4$ ($\sim$3$t_{ff}$)\\
Warm-up phase & 20--200 & $2\times10^{9}$--$8\times10^{12}$ & 40--5 & 120\\
Hot phase & 200--2000 & $8\times10^{12}$--$8\times10^{15}$ & 5--$<$0.1 & 1900\\
\hline
\end{tabular}
\end{center}
\end{table}



\begin{table}
\begin{center}
\caption{Comparison between the protostellar and prestellar hot corino models.}
\footnotesize
\begin{tabular}{ccc}
\hline\hline
  & Protostellar hot corino & Prestellar hot corino \\ 
\hline
Radius ($T$ = 100 K) & 120 AU\footnotemark[1] & $<$10 AU\\
Gas density ($T$ = 100 K) & 10$^7$ cm$^{-3}$\footnotemark[1]& 10$^{10}$ cm$^{-3}$\\
H$_2$CO &2.5(-5)&2.6(-5)\\
CH$_3$OH &1.0(-5)&6.4(-6)\\
HCOOCH$_3$ &3.3(-9)&4.8(-11)\\
CH$_3$OCH$_3$ &6.0(-11)&2.2(-13)\\
HCOOH &3.0(-9)&1.3(-10)\\
\hline
\end{tabular}
\tablecomments{As the protostellar HC model, we adopt the values at the protostellar age of $9.3\times10^4$ yr. As the prestellar HC model, we use the values in the middle stage. Model abundances are represented with respect to H$_2$ molecule at 15 AU in the protostellar HC and at 5 AU in the prestellar HC.}
\tablenotetext{a}{Masunaga \& Inutsuka (2000).}
\end{center}
\end{table}

\clearpage

\setcounter{table}{0}

\appendix
\section{Three Body Association and Collisional Dissociation Reactions}
Table A1 lists the three body association reactions. while Table A2 lists the collisional dissociation reactions in our network.

\begin{table}
\begin{center}
\caption{\footnotesize{Three body association reactions in the reaction network, which are taken from Willacy et al (1998). The rate is calculated as $k=\alpha(T/300)^{\beta}\exp(-\gamma/T)$.}} 
\scriptsize
\begin{tabular}{cccccccccccc}
\hline\hline
       &   &   &   &       & Reaction      &        &   &       & $\alpha$  & $\beta$ & $\gamma$\\
\hline
CH$_3$ & + & H & + & M     & $\rightarrow$ & CH$_4$ & + & M     & 9.62(-31) & -1.80   & 3.23(3)\\
C      & + & O & + & M     & $\rightarrow$ & CO     & + & M     & 2.14(-29) & -3.08   & 2.11(3)\\
CO     & + & O & + & M     & $\rightarrow$ & CO$_2$ & + & M     & 9.56(-34) &  0.00   & 1.06(3)\\
H      & + & CO& + & M     & $\rightarrow$ & HCO    & + & M     & 5.30(-34) &  0.00   & 3.70(2)\\
N      & + & N & + & M     & $\rightarrow$ & N$_2$  & + & M     & 9.53(-34) & -0.50   & 0.00(0)\\
N      & + & H & + & M     & $\rightarrow$ & NH     & + & M     & 3.09(-32) & -0.50   & 0.00(0)\\
NH     & + & H & + & M     & $\rightarrow$ & NH$_2$ & + & M     & 3.18(-33) & -0.50   & 0.00(0)\\
NH$_2$ & + & H & + & M     & $\rightarrow$ & NH$_3$ & + & M     & 1.32(-33) &  0.00   & 4.37(3)\\
N      & + & O & + & M     & $\rightarrow$ & NO     & + & M     & 3.86(-34) &  0.00   & 2.57(3)\\
O      & + & O & + & M     & $\rightarrow$ & O$_2$  & + & M     & 5.25(-35) &  0.00   & 9.02(2)\\
O      & + & H & + & M     & $\rightarrow$ & OH     & + & M     & 6.32(-34) &  0.00   & 1.96(3)\\
OH     & + & H & + & M     & $\rightarrow$ & H$_2$O & + & M     & 6.86(-31) & -2.00   & 0.00(0)\\
S      & + & S & + & M     & $\rightarrow$ & S$_2$  & + & M     & 2.76(-33) & -0.50   & 0.00(0)\\
SO     & + & O & + & M     & $\rightarrow$ & SO$_2$ & + & M     & 1.86(-31) & -0.50   & 0.00(0)\\
H      & + & H & + & H$_2$ & $\rightarrow$ & H$_2$  & + & H$_2$ & 8.72(-33) & -0.60   & 0.00(0)\\
H      & + & H & + & H     & $\rightarrow$ & H$_2$  & + & H     & 1.83(-31) & -1.00   & 0.00(0)\\
\hline
\end{tabular}
\end{center}
\end{table}

\begin{table}
\begin{center}
\caption{\footnotesize{Collisional dissociation reactions in the reaction network. The rate is calculated as $k=\alpha(T/300)^{\beta}\exp(-\gamma/T)$. At the end of the row, W and H indicate that the rate is taken from Willacy et al. (1998) and Harada et al. (2010), respectively. WE also refers Willacy et al., but $\gamma$ is replaced by the difference between the formation enthalpy of the products and the reactants at 300 K. E indicates that the rate is estimated. The rates of the collisional dissociation reactions, which are not listed here, are calculated as $k=10^{-10}\exp(-50000/T)$.}}
\scriptsize
\begin{tabular}{cccccccccccccc}
\hline\hline
        &   &     &   Reaction    &            &   &   &   &   & $\alpha$    & $\beta$ & $\gamma$ & Reference\\
\hline
C$_2$ & + & M &   $\rightarrow$ & C & + & C & + & M & 6.17(-10) & 0.00 & 7.17(4) & WE\\
C$_3$ & + & M &   $\rightarrow$ & C$_2$ & + & C & + & M & 6.17(-10) & 0.00 & 8.83(4) & WE\\
C$_4$ & + & M   & $\rightarrow$ & C$_3$ & + & C & + & M & 1.00(-10) & 0.00 & 6.81(4)& E\\
C$_5$ & + & M   & $\rightarrow$ & C$_4$ & + & C & + & M & 1.00(-10) & 0.00 & 8.52(4)& E\\
CH & + & M &   $\rightarrow$ & C & + & H & + & M & 3.16(-10) & 0.00 & 4.10(4)& WE\\
CH$_2$ & + & M  & $\rightarrow$ & CH & + & H & + & M & 6.64(-9) & 0.00 & 5.12(4)& WE\\
CH$_2$ & + & M   & $\rightarrow$ & C & + & H$_2$ & + & M & 2.16(-10) & 0.00 & 3.97(4)& WE\\
CH$_3$ & + & M   & $\rightarrow$ & CH & + & H$_2$ & + & M & 1.15(-9) & 0.00 & 5.40(4)& WE\\
CH$_3$ & + & M   & $\rightarrow$ & CH$_2$ & + & H & + & M & 1.70(-8) & 0.00 & 5.52(4)& WE\\
CH$_4$ & + & M   & $\rightarrow$ & CH$_3$ & + & H & + & M & 1.20(-6) & 0.00 & 5.28(4)& WE\\
C$_2$H & + & M   & $\rightarrow$ & C$_2$ & + & H & + & M & 3.74(-10) & 0.00 & 6.96(4)& WE\\
C$_2$H$_2$ & + & M   & $\rightarrow$ & C$_2$ & + & H$_2$ & + & M & 7.59(-17) & 0.00 & 7.35(4)& WE\\
C$_2$H$_2$ & + & M   & $\rightarrow$ & C$_2$H & + & H & + & M & 6.97(-8) & 0.00 & 5.63(4)& WE\\
C$_2$H$_3$ & + & M   & $\rightarrow$ & C$_2$H$_2$ & + & H & + & M & 5.07(-20) & -7.50 & 2.29(4)& W\\
C$_2$H$_4$ & + & M   & $\rightarrow$ & C$_2$H$_3$ & + & H & + & M & 1.00(-10) & 0.00 & 5.59(4)& E\\
C$_2$H$_5$ & + & M   & $\rightarrow$ & C$_2$H$_4$ & + & H & + & M & 1.00(-10) & 0.00 & 1.67(4)& E\\
C$_3$H$_4$ & + & M   & $\rightarrow$ & C$_3$H$_3$ & + & H & + & M & 1.66(-7) & 0.00 & 4.47(4)& WE\\
C$_4$H$_2$ & + & M   & $\rightarrow$ & C$_4$H & + & H & + & M & 5.91(-7) & 0.00 & 4.03(4)& W\\
CO & + & M &   $\rightarrow$ & C & + & O & + & M & 1.93(-6) & 0.00 & 1.29(5)& WE\\
CO$_2$ & + & M  & $\rightarrow$ & CO & + & O & + & M & 8.01(-11) & 0.00 & 6.40(4)& WE\\
HCO & + & M   & $\rightarrow$ & CO & + & H & + & M & 2.39(-10) & 0.00 & 8.13(3)& W\\
H$_2$CO & + & M   & $\rightarrow$ & HCO & + & H & + & M & 2.10(-8) & 0.00 & 4.54(4)& WE\\
H$_2$CO & + & M   & $\rightarrow$ & H$_2$ & + & CO & + & M & 3.95(-9) & 0.00 & 1.75(4)& W\\
C$_2$O & + & M     &$\rightarrow$ & C & + & CO & + & M & 1.00(-10) & 0.00 & 3.84(4)& E\\
CS & + & M &   $\rightarrow$ & C & + & S & + & M & 1.00(-10) & 0.00 & 8.58(4)& E\\
HCS & + & M &   $\rightarrow$ & CS & + & H & + & M & 1.00(-10) & 0.00 & 2.39(4)& E\\
H$_2$CS & + & M   & $\rightarrow$ & HCS & + & H & + & M & 1.00(-10) & 0.00 & 4.85(4)& E\\
CN & + & M &   $\rightarrow$ & C & + & N & + & M & 3.32(-10) & 0.00 & 9.07(4)& WE\\
HCN & + & M &   $\rightarrow$ & CN & + & H & + & M & 2.08(-8) & 0.00 & 6.23(4)& WE\\
HNC & + & M &   $\rightarrow$ & CN & + & H & + & M & 1.00(-10) & 0.00 & 5.52(4)& E\\
HNCO & + & M &   $\rightarrow$ & NH & + & CO & + & M & 1.00(-10) & 0.00 & 4.42(4)& E\\
C$_2$N & + & M   & $\rightarrow$ & C & + & CN & + & M & 1.00(-10) & 0.00 & 7.16(4)& E\\
S$_2$ & + & M   & $\rightarrow$ & S & + & S & + & M & 7.95(-11) & 0.00 & 5.12(4)& WE\\
HS & + & M &   $\rightarrow$ & H & + & S & + & M & 1.00(-10) & 0.00 & 4.28(4)& E\\
H$_2$S & + & M   & $\rightarrow$ & HS & + & H & + & M & 7.70(-10) & 0.00 & 4.55(4)& WE\\
SO & + & M &   $\rightarrow$ & S & + & O & + & M & 6.61(-10) & 0.00 & 6.27(4)& WE\\
SO$_2$ & + & M   & $\rightarrow$ & SO & + & O & + & M & 6.60(-9) & 0.00 & 6.63(4)& WE\\
H$_2$S$_2$ & + & M   & $\rightarrow$ & H$_2$S & + & S & + & M & 1.00(-10) & 0.00 & 2.90(4)& E\\
N$_2$ & + & M   & $\rightarrow$ & N & + & N & + & M & 9.21(-5) & -2.50 & 1.14(5)& WE\\
NH & + & M &   $\rightarrow$ & N & + & H & + & M & 4.40(-10) & 0.00 & 3.80(4)& W\\
NH$_2$ & + & M   & $\rightarrow$ & NH & + & H & + & M & 1.00(-36) & 0.50 & 4.91(4)& WE\\
NH$_3$ & + & M   & $\rightarrow$ & NH$_2$ & + & H & + & M & 3.46(-8) & 0.00 & 5.42(4)& WE\\
NH$_3$ & + & M   & $\rightarrow$ & NH & + & H$_2$ & + & M & 1.05(-9) & 0.00 & 5.08(4)& WE\\
NS & + & M &   $\rightarrow$ & N & + & S & + & M & 1.00(-10) & 0.00 & 5.85(4)& E\\
NO & + & M &   $\rightarrow$ & N & + & O & + & M & 4.10(-9) & 0.00 & 7.60(4)& WE\\
HNO & + & M     &$\rightarrow$ & H & + & NO & + & M & 1.00(-10) & 0.00 & 2.36(4)& E\\
N$_2$O & + & M   & $\rightarrow$ & N & + & NO & + & M & 1.00(-10) & 0.00 & 5.79(4)& E\\
SiH & + & M &   $\rightarrow$ & Si & + & H & + & M & 1.00(-10) & 0.00 & 3.51(4)& E\\
SiH$_2$ & + & M   & $\rightarrow$ & SiH & + & H & + & M & 7.60(-8) & -1.80 & 3.87(4)& WE\\
SiH$_2$ & + & M   & $\rightarrow$ & Si & + & H$_2$ & + & M & 7.60(-8) & -1.76 & 2.13(4)& WE\\
SiH$_3$ & + & M   & $\rightarrow$ & SiH & + & H$_2$ & + & M & 4.98(-10) & 0.00 & 2.12(4)& WE\\
SiH$_4$ & + & M   & $\rightarrow$ & SiH$_3$ & + & H & + & M & 5.00(-11) & 0.00 & 4.62(4)& WE\\
SiH$_4$ & + & M   & $\rightarrow$ & SiH$_2$ & + & H$_2$ & + & M & 5.00(-11) & 0.00 & 2.90(4)& WE\\
\hline
\end{tabular}
\end{center}
\end{table}

\setcounter{table}{1}

\begin{table}
\begin{center}
\caption{continue}
\scriptsize
\begin{tabular}{ccccccccccccc}
\hline\hline
    &   &     &   Reaction    &            &   &   &   &   & $\alpha$    & $\beta$ & $\gamma$ & Reference\\
\hline
SiC & + & M &   $\rightarrow$ & Si & + & C & + & M & 1.00(-10) & 0.00 & 5.38(4)& E\\
SiS & + & M &   $\rightarrow$ & Si & + & S & + & M & 1.00(-10) & 0.00 & 7.47(4)& E\\
SiN & + & M &   $\rightarrow$ & Si & + & N & + & M & 1.00(-10) & 0.00 & 6.62(4)& E\\
SiO & + & M &   $\rightarrow$ & Si & + & O & + & M & 1.00(-10) & 0.00 & 9.62(4)& E\\
SiO$_2$ & + & M   & $\rightarrow$ & SIO & + & O & + & M & 1.00(-10) & 0.00 & 5.46(4)& E\\
SiC$_2$ & + & M   & $\rightarrow$ & SiC & + & C & + & M & 1.00(-10) & 0.00 & 9.95(4)& E\\
OH & + & M &   $\rightarrow$ & O & + & H & + & M & 4.00(-9) & 0.00 & 5.15(4)& WE\\
H$_2$O & + & M   & $\rightarrow$ & OH & + & H & + & M & 5.80(-9) & 0.00 & 6.00(4)& WE\\
OCS & + & M &   $\rightarrow$ & CO & + & S & + & M & 4.82(-10) & 0.00 & 3.71(4)& WE\\
HCl & + & M &   $\rightarrow$ & H & + & CL & + & M & 1.00(-10) & 0.00 & 5.19(4)& E\\
CCl & + & M &   $\rightarrow$ & C & + & Cl & + & M & 1.00(-10) & 0.00 & 4.04(4)& E\\
ClO & + & M &   $\rightarrow$ & Cl & + & O & + & M & 1.00(-10) & 0.00 & 3.24(4)& E\\
CP & + & M &   $\rightarrow$ & C & + & P & + & M & 1.00(-10) & 0.00 & 7.02(4)& E\\
HCP & + & M     &$\rightarrow$ & H & + & CP & + & M & 1.00(-10) & 0.00 & 6.23(4)& E\\
CH$_2$PH & + & M   & $\rightarrow$ & CH$_2$ & + & PH & + & M & 1.00(-10) & 0.00 & 1.11(5)& E\\
MgH & + & M   & $\rightarrow$ & Mg & + & H & + & M & 1.00(-10) & 0.00 & 2.36(4)& E\\
NaH & + & M     &$\rightarrow$ & Na & + & H & + & M & 1.00(-10) & 0.00 & 2.42(4)& E\\
PH & + & M      &$\rightarrow$ & P & + & H & + & M & 1.00(-10) & 0.00 & 3.38(4)& E\\
PH$_2$ & + & M &   $\rightarrow$ & PH & + & H & + & M & 1.00(-10) & 0.00 & 4.13(4)& E\\
PN & + & M     &$\rightarrow$ & P & + & N & + & M & 1.00(-10) & 0.00 & 8.23(4)& E\\
PO & + & M    &$\rightarrow$ & P & + & O & + & M & 1.00(-10) & 0.00 & 7.09(4)& E\\
HPO & + & M   &  $\rightarrow$ & H & + & PO & + & M & 1.00(-10) & 0.00 & 3.02(4)& E\\
O$_2$ & + & M   & $\rightarrow$ & O & + & O & + & M & 5.16(-10) & 0.00 & 6.00(4)& WE\\
O$_2$H & + & M &   $\rightarrow$ & O$_2$ & + & H & + & M & 1.00(-10) & 0.00 & 2.47(4)& E\\
H$_2$O$_2$ & + & M   & $\rightarrow$ & H$_2$O & + & O & + & M & 1.00(-10) & 0.00 & 1.73(4)& E\\
O$_3$ & + & M &   $\rightarrow$ & O$_2$ & + & O & + & M & 1.00(-10) & 0.00 & 1.28(4)& E\\
CH$_3$N & + & M   & $\rightarrow$ & CH$_3$ & + & N & + & M & 1.00(-10) & 0.00 & 6.34(4)& E\\
CH$_3$NH$_2$ & + & M &   $\rightarrow$ & CH$_3$ & + & NH$_2$ & + & M & 1.00(-10) & 0.00 & 4.26(4)& E\\
CH$_2$CN & + & M   & $\rightarrow$ & C$_2$N & + & H$_2$ & + & M & 1.00(-10) & 0.00 & 6.00(4)& E\\
CH$_3$CN & + & M   & $\rightarrow$ & CH$_3$ & + & CN & + & M & 1.00(-10) & 0.00 & 6.21(4)& E\\
HCOOH & + & M   & $\rightarrow$ & HCO & + & OH & + & M & 1.00(-10) & 0.00 & 5.55(4)& E\\
CH$_2$CO & + & M   & $\rightarrow$ & CH$_2$ & + & CO & + & M & 5.97(-9) & 0.00 & 3.86(4)& WE\\
CH$_3$OH & + & M &   $\rightarrow$ & CH$_3$ & + & OH & + & M & 3.32(-7) & 0.00 & 4.64(4)& WE\\
CH$_3$CHO & + & M   & $\rightarrow$ & CH$_3$ & + & HCO & + & M & 1.00(-10) & 0.00 & 4.20(4)& E\\
H$_2$ & + & e$^-$ & $\rightarrow$ & H & + & H & + & e$^-$ & 2.00(-9) & 0.50 & 1.20(5)& H\\
H$_2$ & + & H & $\rightarrow$ & H & + & H & + & H & 1.00(-10) & 0.00 & 5.20(4)& H\\
H$_2$ & + & He & $\rightarrow$ & H & + & H & + & He & 1.00(-11) & 0.00 & 5.20(4)& H\\
H$_2$ & + & H$_2$ & $\rightarrow$ & H & + & H & + & H$_2$ & 1.30(-11) & 0.00 & 5.20(4)& H\\
\hline

\end{tabular}
\end{center}
\end{table}


\begin{thebibliography}{}
\bibitem[Ag\'{u}ndez et al. (2008)]{agundez08} Ag\'{u}ndez, M., Cernicharo, J., Gu\'{e}lin, M., Gerin, M., McCarthy, M. C., \& Thaddeus, P. 2008, \aap, 478, L19
\bibitem[Aikawa et al. (2008)]{aikawa08} Aikawa, Y., Wakelam, V., Garrod, R. T., \& Herbst, E. 2008, \apj, 674, 984
\bibitem[Bate (2010)]{bate10} Bate, M. R., 2010, MNRAS, 404, L79
\bibitem[Bate (2011)]{bate11} Bate, M. R., 2011, MNRAS, 417, 2036
\bibitem[Bohlin et al. (1978)]{bohlin78} Bohlin, R. C., Savage, B. D, \& Drake, J. F. 1978, \apj, 224, 132
\bibitem[Bottinelli et al. (2004)]{bottinelli04} Bottinelli, S., et al. 2004. \apj, 615, 354
\bibitem[Br\"{u}nken et al. (2007)]{brunken07} Br\"{u}nken, S., Gupta, H., Gottlieb, C. A., McCarthy, M. C., \& Thaddeus, P. 2007, \apj, 664, L43
\bibitem[Charnley et al. (1992)]{charnley92} Charnley, S. B., Tielens, A. G. G. M., \& Millar, T. J. 1992, \apj, 399, L71  
\bibitem[Chen et al. (2010)]{chen10} Chen, X., Arce, H. G., Zhang, Q., Bourke, T. L., Launhardt, R., Schmalzl, M., \& Henning, T. 2010, \apj, 715, 1344
\bibitem[Caselli et al. (1998)]{caselli98} Caselli, P., Hasegawa, T. I., \& Herbst, E. 1998, \apj, 495, 309
\bibitem[Cardelli et al. (1989)]{cardelli89} Cardelli, J. A., Clayton, G. C., \& Mathis. J. S. 1989, \apj, 345, 245
\bibitem[Cazaux et al. (2003)]{cazaux03} Cazaux, S., et al. 2003, \apj, 593, L51
\bibitem[Chandler et al. (2005)]{chandler05} Chandler, C. J., Brogan, C. L., Shirley, Y. L., \& Loinard, L. 2005, \apj, 632, 371
\bibitem[Commer\c{c}on et al. (2011)]{commercon11} Commer\c{c}on, B., Audit, E., Chabrier, G., \& Chi\`{e}ze, J. -P. 2011, \aap, 530, 13
\bibitem[Commer\c{c}on et al. (2010)]{commercon10} Commer\c{c}on, B., Hennebelle, P., Audit, E., Chabrier, G., \& Teyssier, R. 2010, \aap, 510, L3
\bibitem[Dalgarno (2006)]{dalgarno06} Dalgarno, A. 2006, Proc. Natl Acad. Sci., 103, 12269
\bibitem[Draine \& Sutin (1987)]{draine87} Draine, B. T., \& Sutin, B., 1987, \apj, 320, 803
\bibitem[Dunham et al. (2011)]{dunham11} Dunham, M, M., Chen, X., Arce, H. G., Bourke, T. L., Schnee, S., \& Enoch, M, L. 2011, \apj, 742, 1
\bibitem[Enoch et al. (2010)]{enoch10} Enoch, M. L., Lee, J., Harvey, P., Dunham, M. M., \& Schnee, S. 2010, \apj, 722, L33
\bibitem[Garrod \& Herbst (2006)]{garrod06} Garrod, R. T., \& Herbst, E. 2006, \aap, 457, 927
\bibitem[Garrod et al. (2007)]{garrod07} Garrod, R. T., Wakelam, V., \& Herbst, E. 2007, \aap, 467, 1103
\bibitem[Garrod et al. (2008)]{garrod08} Garrod, R. T., Weaver, S. L. W., \& Herbst, E. 2008, \apj, 682, 283
\bibitem[Geppert et al. (2006)]{geppert06} Geppert, W. D., et al. 2006, Faraday Discuss. 133, Chemical Evolution of the Universe (Cambridge: RSC Pub.), 177
\bibitem[Graedel et al. (1982)]{graedel82} Graedel, T. E., Langer, W. D., \& Frerking, M. A. 1982, \apjs, 48, 321
\bibitem[Hamberg et al. (2010)]{hamberg10} Hamberg, M., et al. 2010, \aap, 514, 83
\bibitem[Harada et al. (2010)]{harada10} Harada, N., Herbst, R., \& Wakelam, V. 2010, \apj, 721, 1570
\bibitem[Hasegawa \& Herbst (1993)]{hasegawa93} Hasegawa, T. I., \& Herbst, E. 1993, MNRAS, 261, 83
\bibitem[Hasegawa et al. (1992)]{hasegawa92} Hasegawa, T. I., Herbst, E., \& Leung, C. M. 1992, \apjs, 83, 167
\bibitem[Hassel et al. (2008)]{hassel08} Hassel, G. E., Herbst, E., \& Garrod, R. T. 2008, \apj, 681, 1385
\bibitem[Herbst \& van Dishoeck (2009)]{herbst09} Herbst, E, \& van Dishoeck, E. F. 2009, ARAA, 47, 427
\bibitem[Hersant et al. (2009)]{hersant09} Hersant, F., Wakelam, V., Dutrey, A., Guilloteau, S., \& Herbst, E. 2009, \aap,
493, L49
\bibitem[Hildebrand (1983)]{hildebrand83} Hildebrand, R. H. 1983, QJRAS, 24, 267
\bibitem[Hollenbach \& Mckee (1979)]{hollenbach79} Hollenbach, D., \& McKee, C. F. 1979, \apjs, 41, 555
\bibitem[Kuan et al. (2004)]{kuan04} Kuan, Y.-J., et al. 2004, \apj, 616, L27
\bibitem[Langer \& Penzias (1993)]{langer93} Langer, W. D., \& Penzias, A. A. 1993, \apj, 408, 539
\bibitem[Larson (1969)]{larson69} Larson, R. B., 1969, MNRAS, 145, 271
\bibitem[Levermore \& Pomraning (1981)]{levermore81} Levermore, C. D., \& Pomraning, G. C. 1981, \apj, 248, 321
\bibitem[Lucas \& Liszt (1998)]{lucas98} Lucas, R., \& Liszt, H. 1998, \aap, 337, 246
\bibitem[Machida et al. (2008)]{machida08} Machida, M. N., Inutsuka, S., \& Matsumoto, T. 2008, \apj, 676, 1088
\bibitem[Machida et al. (2010)]{machida10} Machida, M. N., Inutsuka, S., \& Matsumoto, T. 2010, \apj, 724, 1006
\bibitem[Machida et al. (2011)]{machida11} Machida, M. N., \& Matsumoto, T. 2011, MNRAS, 413, 2767
\bibitem[Maret et al. (2005)]{maret05} Maret, S., et al. 2005, \aap, 442, 527
\bibitem[Masunaga \& Inutsuka (2000)]{masunaga00} Masunaga, H., \& Inutsuka, S. 2000, \apj, 531, 350
\bibitem[Masunaga et al. (1998)]{masunaga98} Masunaga, H., Miyama, S. M., \& Inutsuka, S. 1998, \apj, 495, 346
\bibitem[Matsumoto \& Hanawa (2003)]{matsumoto03} Matsumoto, T., \& Hanawa, T. 2003, \apj, 595, 913
\bibitem[Omukai (2007)]{omukai07} Omukai K., 2007, PASJ, 59, 589
\bibitem[Pineda et al. (2011)]{pineda11} Pineda, J. E., et al. 2011, \apj, 743, 201
\bibitem[Sakai et al. (2009)]{sakai09} Sakai, N., Sakai, T., Hirota, T., Burton, M., \& Yamamoto, S. 2009, \apj, 697, 769
\bibitem[Sakai et al. (2008)]{sakai08} Sakai, N., Sakai, T., Hirota, T., \& Yamamoto, S. 2008, \apj, 672, 371
\bibitem[Sakai et al. (2007)]{sakai07} Sakai, N., Sakai, T., Osamura, Y., \& Yamamoto, S. 2007, \apj, 667, L65
\bibitem[Saigo et al. (2008)]{saigo08} Saigo, K., Tomisaka, K., \& Matsumoto, T. 2008, \apj, 674, 997
\bibitem[Saigo \& Tomisaka (2011)]{saigo11} Saigo, K., \& Tomisaka, K. 2011, \apj, 728, 78
\bibitem[Semenov et al. (2010)]{semenov10} Semenov, D., et al. 2010, \aap, 522, 42
\bibitem[Tielens (2005)]{tielens05} Tielens, A. G. G. M. 2005, The Physics and Chemistry of the Interstellar Medium (Cambridge: Cambridge Univ. Press), 219
\bibitem[Tomida (2012)]{tomida12a} Tomida, K. 2012, PhD thesis, The Graduate University for Advanced Studies (SOKENDAI), Osawa, Mitaka, Tokyo 181-8588, Japan
\bibitem[Tomida et al. (2010a)]{tomida10a} Tomida, K., Tomisaka, K., Matsumoto, T., Ohsuga, K., Machida, M. N., \& Saigo, K. 2010a, \apj, 714, L58
\bibitem[Tomida et al. (2012)]{tomida12b} Tomida, K., Tomisaka, K., Matsumoto, T., Hori, Y., Okuzumi, S., Machida, M. N., \& Saigo, K. 2012, arXiv:1206.3567
\bibitem[Tomida et al. (2010b)]{tomida10b} Tomida, K., Machida, M. N., Saigo, K., Tomisaka, K., \& Matsumoto, T. 2010b, \apj, 725, L239
\bibitem[Tomisaka (2002)]{tomisaka02} Tomisaka, K. 2002, \apj, 575, 306
\bibitem[Tomisaka \& Tomida (2011)]{tomisaka11} Tomisaka, K. \& Tomida, K. 2011, PASJ, 63, 1151
\bibitem[Umebayashi (1983)]{umebayashi83} Umebayashi, T. 1983, Prog. Theor. Phys., 69, 480
\bibitem[Umebayashi \& Nakano (1980)]{umebayashi80} Umebayashi, T., \& Nakano, T. 1980, PASJ, 32, 405
\bibitem[Umebayashi \& Nakano (1981)]{umebayashi81} Umebayashi, T., \& Nakano, T. 1981, PASJ, 33, 617
\bibitem[Umebayashi \& Nakano (2009)]{umebayashi09} Umebayashi, T., \& Nakano, T. 2009, \apj, 690, 69
\bibitem[van Weeren et al. (2009)]{weeren09} van Weeren, R. J., Brinch, C., \& Hogerheijde, M. R. 2009, \aap, 497, 773
\bibitem[Visser et al. (2011)]{visser11} Visser, R., Doty, S. D., \& van Dishoeck, E. F. 2011, \aap, 534, 132
\bibitem[Wakelam et al. (2012)]{wakelam12} Wakelam, V., et al. 2012, ApJS, 199, 21 
\bibitem[Walker et al. (1988)]{walker88} Walker, C. K., Lada, C. J., Young, E. T., \& Margulis, M. 1988, \apj, 332, 335
\bibitem[Watson \& Salpeter (1972)]{watson72} Watson, W. D., \& Salpeter, E. E. 1972, \apj, 174, 321 
\bibitem[Whitehouse \& Bate (2006)]{white06} Whitehouse, S. C., \& Bate, M. R. 2006, MNRAS, 367, 32
\bibitem[Willacy et al. (1998)]{willacy98} Willacy, K., Klahr, H. H., Millar, T. J., \& Henning, Th. 1998, \aap, 338, 995
\bibitem[Yorke \& Bodenheimer (1999)]{yorke99} Yorke, H. W., \& Bodenheimer, P. 1999, \apj, 525, 330 
  
\end{thebibliography}
\end{document}